\documentclass{aa}  
\usepackage{graphicx}
\usepackage{txfonts}
\usepackage{amsmath}
\usepackage{amssymb}
\usepackage{mathbbol}
 \usepackage{mathrsfs}
\usepackage{dsfont}
\usepackage{amsfonts}
\usepackage{color}
\usepackage{amsopn}
\usepackage{bm}
\usepackage{graphicx,longtable,times,threeparttable,tablefootnote, multirow}
\usepackage{longtable,times,threeparttable,multirow}
\usepackage{float}

\DeclareGraphicsExtensions{.eps,.eps.gz,.epsi, .jpg, .png}

\newcommand{\teff}{\ensuremath{T_{\mathrm {eff}}\,}}
\newcommand{\logg}{\ensuremath{{\mathrm {log}\, } g\,}}
\newcommand{\feh}{\ensuremath{[{\mathrm {Fe/H}}]\,}}
\newcommand{\sih}{\ensuremath{[{\mathrm {Si/H}}]\,}}
\newcommand{\vsini}{\ensuremath{v{\mathrm {sin}}i\,}}
\newcommand{\vmic}{\ensuremath{v_{\mathrm {mic}}\,}}
\newcommand{\mk}{\ensuremath{M_{K{\mathrm s}}\,}}
\newcommand{\msun}{\ensuremath{M_\odot\,}}

\newcommand{\payne}{{\sc Payne}\,}
\newcommand{\hotpayne}{{\sc HotPayne}\,}
\newcommand{\he}{\ensuremath{{N_{\mathrm {He}}/N_{\mathrm {tot}}}\,}}

\begin{document} 
   \title{Stellar labels for hot stars from low-resolution spectra -- I. the \emph{HotPayne} method and results for $330,000$ stars from LAMOST DR6}
   \author{Maosheng Xiang,
          \inst{1}\thanks{mxiang@mpia.de}
          Hans-Walter Rix, 
          \inst{1} 
          Yuan-Sen Ting,
          \inst{2,3,4,5}\thanks{Hubble fellow}
          Rolf-Peter Kudritzki, 
          \inst{6, 7}  
          Charlie Conroy,
          \inst{8}
          Eleonora Zari,
          \inst{1}
          Jian-Rong Shi,
          \inst{9,10}
          Norbert Przybilla, 
          \inst{11}   
          Maria Ramirez-Tannus,
          \inst{1}
          Andrew Tkachenko, 
          \inst{12}
          Sarah Gebruers,
          \inst{12}
          Xiao-Wei Liu
          \inst{13}
          }
\institute{Max-Planck Institute for Astronomy, K\"onigstuhl 17, D-69117 Heidelberg, Germany  
\and
   Institute for Advanced Study, Princeton, NJ 08540, USA
\and
   Department of Astrophysical Sciences, Princeton University, Princeton, NJ 08544, USA
\and
   Observatories of the Carnegie Institution of Washington, 813 Santa
Barbara Street, Pasadena, CA 91101, USA
\and
   Research School of Astronomy \& Astrophysics, Australian National University, Canberra, ACT 2611, Australia
\and
   LMU M\"unchen, Universit\"atatssternwarte, Scheinerstr. 1, 81679 M\"unchen, Germany
\and
   Institute for Astronomy, University of Hawaii at Manoa, 2680 Woodlawn Drive, Honolulu, HI 96822, USA 
\and
   Center for Astrophysics, Harvard \& Smithsonian, Cambridge, MA 02138, USA
\and
  National Astronomical Observatories, Chinese Academy of Sciences, Beijing 100012, P. R. China
\and
   University of Chinese Academy of Sciences, Beijing 100049, P. R. China
\and
   Institut f\"ur Astro- und Teilchenphysik, Universit\"at Innsbruck,
              Technikerstrasse 25, 6020 Innsbruck, Austria   
\and
   Institute of Astronomy, KU Leuven, Celestijnenlaan 200D, 3001 Leuven, Belgium
\and   
   South-Western Institute for Astronomy Research, Yunnan University, Kunming 650500, P. R. China 
             }
  \date{Submitted to A\&A on June 17, 2021}
  \abstract
 {We set out to determine stellar labels from low-resolution survey spectra of \emph{hot stars}, specifically OBA stars with $\teff\gtrsim 7500$~K. This fills a gap in the scientific analysis of large spectroscopic stellar surveys such as LAMOST, which offers spectra for millions of stars at $R\sim 1800$ and covering $3800\AA\le \lambda \le 9000\AA$. We first explore the theoretical information content of such spectra for determining stellar labels, via the Cram\'er-Rao bound. We show that in the limit of perfect model spectra and observed spectra with S/N$\sim 50-100$, precise estimates are possible for a wide range of stellar labels: not only the effective temperature \teff, surface gravity \logg, and projected rotation velocity \vsini, but also the micro-turbulence velocity \vmic, Helium abundance \he and the elemental abundances [C/H], [N/H], [O/H], [Si/H], [S/H], and [Fe/H]. Our analysis illustrates that the temperature regime of $\teff\sim 9500$~K is challenging, as the dominant Balmer and Paschen line strength vary little with \teff. We implement the simultaneous fitting of these 11 stellar labels to LAMOST hot-star spectra using the \payne approach, drawing on Kurucz's ATLAS12/SYNTHE LTE spectra as the underlying models. We then obtain stellar parameter estimates for a sample of about 330,000 hot stars with LAMOST spectra, an increase by about two orders of magnitude in sample size. Among them, about 260,000 have $S/N>5$ Gaia parallaxes, and their luminosities imply that $\gtrsim 95\%$ among them are luminous star, mostly on the main sequence; the rest reflects lower luminosity evolved stars, such as hot subdwarfs and white dwarfs. We show that the fidelity of the results, particularly for the abundance estimates, is limited by the systematics of the underlying models, as they do not account for NLTE effects. Finally, we show the detailed distribution of $v\sin{i}$ of stars with $8000-15,000$~K, illustrating that it extends to a sharp cut-off at the critical rotation velocity, $v_{\rm crit}$, across a wide range of temperatures. }
 
\authorrunning{Maosheng Xiang et al.}
\titlerunning{Stellar label determination for hot stars from LAMOST low-resolution spectra }

\maketitle 

%
\section{Introduction}\label{sec:introduction}
Hot stars, here meaning stars with OBA spectral types or $T_{\rm eff}\gtrsim7500$\,K, are interesting objects for a number of reasons. They are tracers of young stellar population in our Galaxy; they are intriguing laboratories to test stellar structure and evolution scenarios; and they can be progenitors of and companions to black holes and supernovae. Recently, these stars have garnered more attention thanks to large sky low-resolution ($R\sim2000$) spectroscopic surveys, such as the LAMOST Galactic surveys \citep{Deng2012, Zhao2012, LiuX2014, Yuan2015} and SDSS-V Milky Way Mapper \citep{Kollmeier2017,Zari2021}. LAMOST has collected $R\sim1800$ optical spectra for thousands of stars identified as OB stars \citep{LiuZ2019}, and hundreds of thousands of A-type stars\footnote{lamost.dr6.org} \citep{Luo2015}. The SDSS-V will yield spectra for many more OB stars in the next few years \citep{Zari2021}. 
Due to their high effective temperature, the spectra of hot stars show much weaker spectral features than the cool stars. As a result, the latter has been the \emph{analysis focus} of most Galactic archaeology surveys to date. Stellar parameter pipelines for large spectroscopic surveys have been mainly focused on FGK stars \citep[e.g.][]{Lee2008, Luo2015, Xiang2015, Xiang2019, Garcia_Perez2016, Casey+2017, Ahumada2020, Buder2020, Steinmetz2020}. Beside the easier measurements, the detailed abundances [X/H] of cool stars have been of particular interest, as they are closer to the birth-material composition of the stars, and may enable ``chemical tagging''. Furthermore, it has been shown that even low-resolution spectra can yield useful stellar labels, such as $\teff$, $\logg$ and abundance [X/H] for $>$15 elements \citep{Ting+2017a,Ting+2017b,Xiang2019}

For hot stars, the surface abundances can be affected by stellar internal mixing process, making them deviate significantly from the solar abundance scale \citep[e.g.][]{Przybilla2010, Ekstrom2012, Maeder2014, Martins2015, Pedersen2018, Aerts2021} and rendering them less useful for inferences in Galactic archaeology. On the other hand, these mixing processes can make hot stars powerful probes of stellar evolution physics. Particularly, it has long been known that ``chemically peculiar'' stars are common among hot stars \citep[e.g.][]{Renson2009, Gray2016, Qin2019, Chojnowski2020, Xiang2020, Paunzen2021}, which provide good constraints on these stellar element transport processes \citep[e.g.][]{Talon2006, Vick2010, Michaud2015}. Despite their interesting prospects, the parameters and detailed abundances of hot stars were only derived for relatively small samples, usually no larger than $\sim\mathrm{O}(10^3)$ stars \citep[e.g.][]{Przybilla2008, Nieva2012, Nieva2014, Simon-Diaz2014, Kudritzki2016, Berger2018, Holgado2018}.

The lack of systematic analysis of hot-star spectra calls for flexible and sophisticated spectral modelling to deliver astrophysical labels for hot stars from the vast sets of low-resolution spectra collected by large sky surveys. This is particularly important for the LAMOST Galactic survey, as it has collected low-resolution ($R\sim1800$) spectra for hundreds of thousands of OBA stars. Yet, a comprehensive and quantitative determination of stellar parameters and abundances for these LAMOST hot-star spectra is largely non-existent, and this is the aspect that we aim to contribute with this study.

In this work, we will demonstrate that $R\sim1800$ spectra contain a wealth of astrophysical information for hot stars. We apply the \payne \citep{Ting+2019} for an ``industrial scale'' determination of stellar labels from a massive set of  hot-star spectra of LAMOST. The \payne is a spectral model fitting tool, with the ability to determine simultaneously many labels from the full spectra at modest computational expense. This is owed to the flexible neural-net based spectral interpolation algorithm that can precisely interpolate spectra in high dimensions \citep{Ting+2019}. Fitting the entire spectrum allows for an optimal use of its information content, and fitting {\it all} pertinent labels simultaneously ensures that label co-variances, due to line blending, are correctly accounted for. We fit a large number (11) of labels to LAMOST spectra of probable hot stars --- effective temperature \teff, surface gravity \logg, micro-turbulence velocity \vmic, projected rotation velocity \vsini, Helium abundance \he, elemental abundances [C/H], [N/H], [O/H], [Si/H], [S/H], and [Fe/H]. We dub our modelling \hotpayne , as the original \payne  approach has been mostly applied to FGK stars; hot star spectra are sufficiently different so that this spectral regime requires non-trivial modifications and testing, as we will show.

We adopt the Kurucz LTE spectra \citep{Kurucz1970, Kurucz1993, Kurucz2005} as our underlying model spectra to train the neural network spectral model of the \hotpayne. As NLTE effects and stellar winds play important roles for hot-star spectroscopy \citep[e.g.][]{Auer1972, Kudritzki1976, Kudritzki1979, Hubeny1995, Kudritzki1998, Kudritzki2000, Lanz2007, Przybilla2008, Przybilla2010, Hainich2019}, our resultant label estimates, based on the Kurucz LTE model spectra, are therefore inevitably affected by non-negligible systematic errors. We will address such systematics by comparing our results with literature values from high-resolution NLTE spectroscopic analysis for a sample of reference stars. 
We emphasize from the outset, that the approach presented here deserves to be followed-up with label determinations based on NLTE model spectra in order to obtain accurate label estimates, in particular of elemental abundances for these hot stars, which we will perform in a future study. Nonetheless, the internal precision for our LTE-based abundance estimates appear to be promising for some interesting applications as presented in this study.  

The paper is organised as follows. In Section~\ref{sec:LAMOSTspectra} we introduce the LAMOST spectra for the sample of candidate hot stars. In Section~\ref{sec:crbound} we explore the information content of such low-resolution spectra of hot stars. In Section~\ref{sec:getting_stellar_labels} we derive stellar labels from the $R\sim1800$ LAMOST spectra for $>330,000$ hot stars. In Section~\ref{sec:results_and_discussion} we present and discuss these results, followed by a summary in Section~\ref{sec:summary}.

\section{LAMOST spectra of candidate hot stars}\label{sec:LAMOSTspectra}
We compile a sample of candidate hot stars with low-resolution ($R\sim1800$) optical ($\lambda\sim$3800--9000) spectra from the sixth data release (DR6)\footnote{dr6.lamost.org} of the LAMOST Galactic surveys \citep{Deng2012, Zhao2012, LiuX2014}; we refer to this initial list as candidates, as our subsequent spectral analysis reveals that some of them have low temperatures of $\teff<7000$~K. The LAMOST Galactic surveys target millions of stars in the full color-magnitude diagram (CMD) down to about 17.8\,mag in SDSS $r$-band \citep{Carlin2012, LiuX2014, Yuan2015}. As a key component of the LAMOST Galactic surveys, the LAMOST spectroscopic survey at the Galactic anti-center (LSS-GAC) uniformly samples stars in the ($g$--$r$, $r$) and ($r-i$, $r$) space. This uniform selection implies that stars in the low density regimes of the CMD (e.g. OB stars) are targeted with higher completeness \citep{LiuX2014, Yuan2015, Xiang2017b, Chen2018} which results in a large sample of hot star spectra with well-characterized completeness, located mostly in the Galactic anti-center direction. The LAMOST DR6 released 9,911,337 spectra, and each has a spectral classification with the LAMOST 1D pipeline \citep[e.g.][]{Luo2015}. Among the 9,911,337 spectra, 9,231,057 are classified as stars. 

\begin{figure*}[hbt!]
\centering
\includegraphics[width=1.0\textwidth]{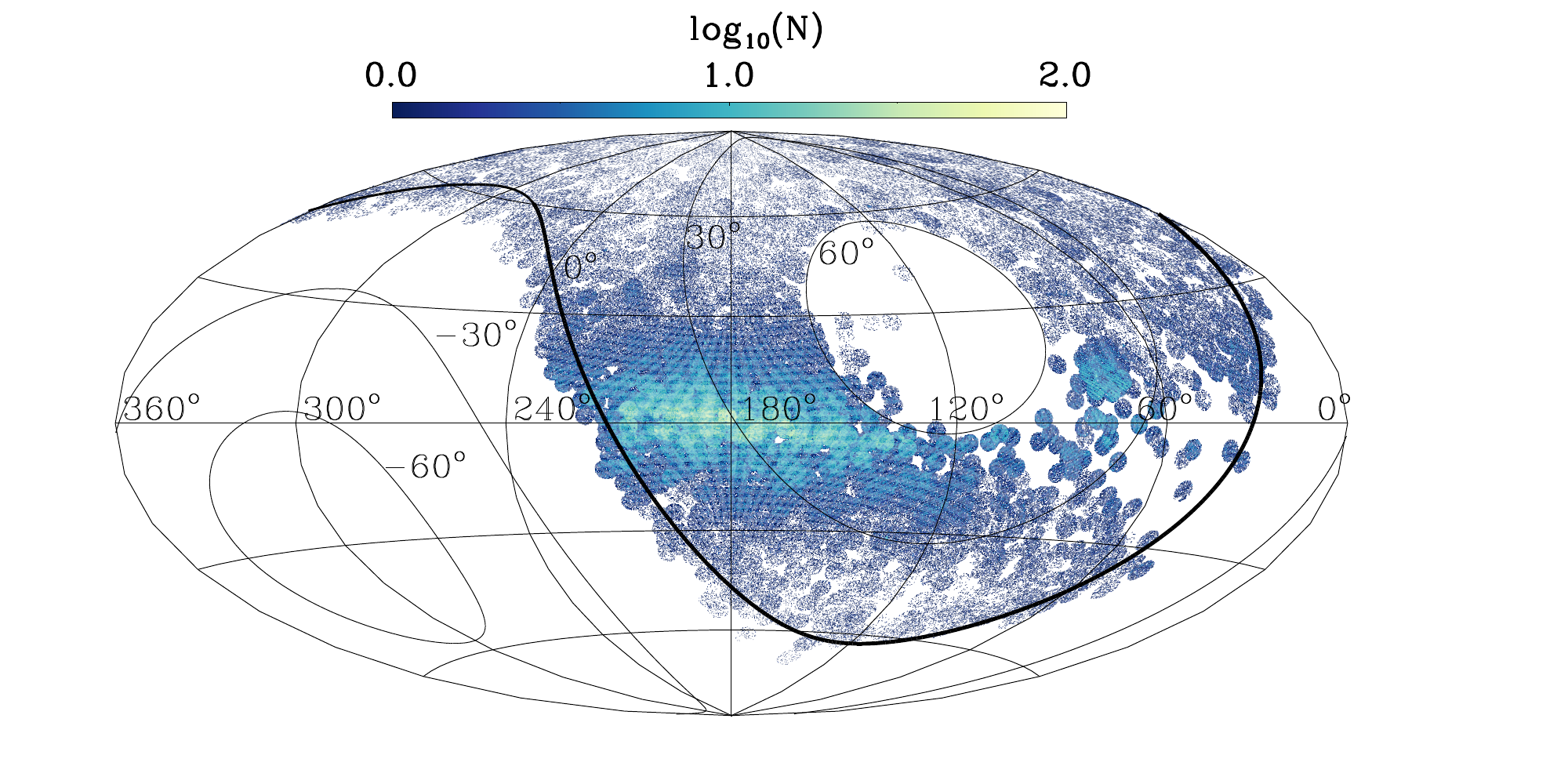}
\caption{On-sky distribution of our LAMOST DR6 hot star candidates in the Galactic coordinates ($l$, $b$). The map is centered on the Galactic anti-center ($l=180^\circ$, $b=0^\circ$). The thick line delineates the celestial equator (Decl. = 0$^\circ$), and lines of constant declination, Decl. = $\pm30^\circ$ and Decl. = $\pm60^\circ$, are also shown. Colors indicate the logarithmic number of stars in each ($l$, $b$) cell of $0.2^\circ\times0.2^\circ$. The distribution reflects both the LAMOST survey area and the candidate stars' concentration to the Galactic plane, which is expected for young stars.}
\label{fig:fig1}%
\end{figure*}

\begin{figure}[hbt!]
\centering
\includegraphics[width=0.5\textwidth]{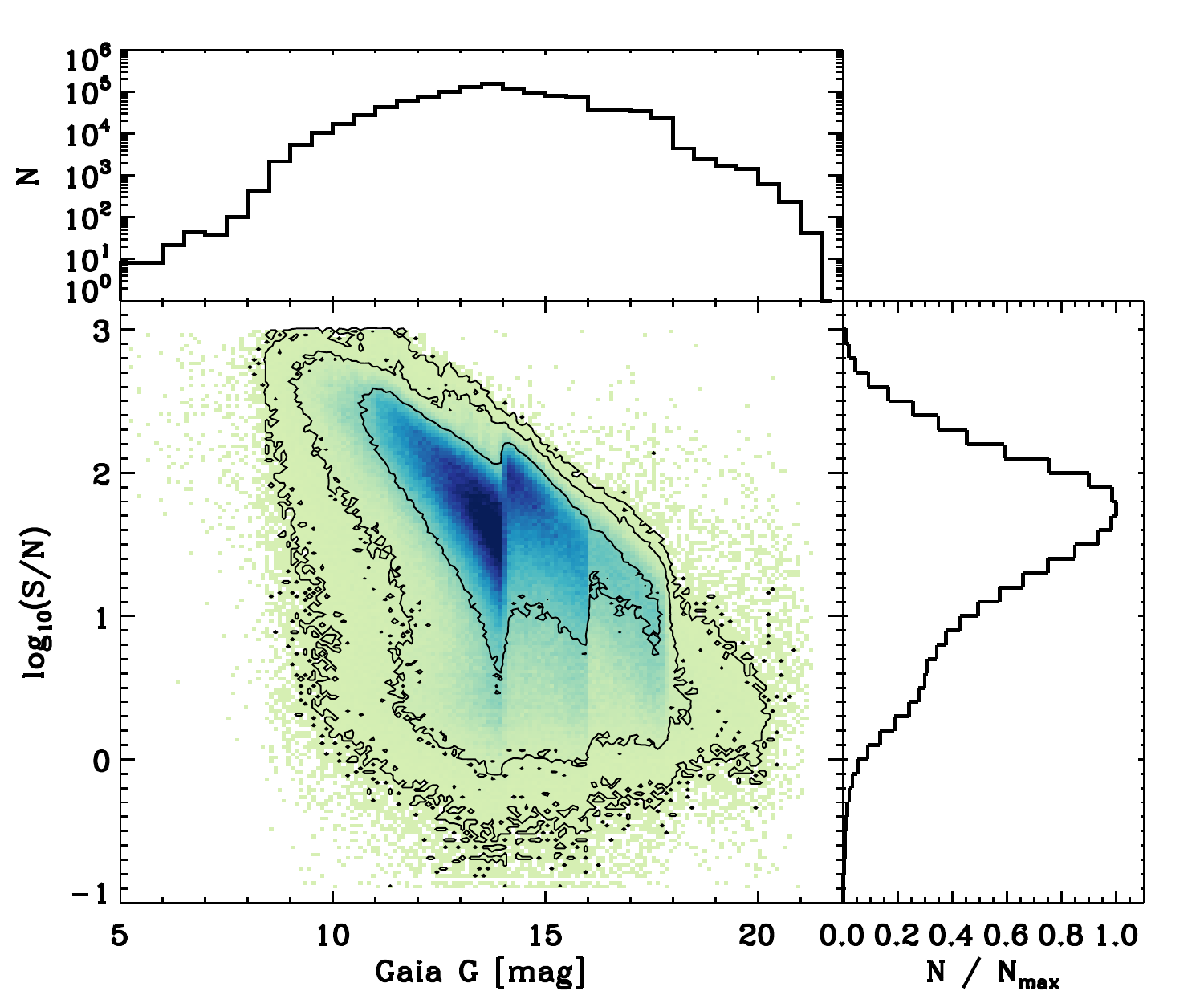}
\caption{Distribution of our LAMOST DR6 hot star candidates in the {\it Gaia} $G$-band magnitude and LAMOST spectral S/N plane. The $G-$band magnitudes cover a dynamic range of nearly 15 magnitudes, and most of them are between $G\sim10$ and 18\,mag. The spectral S/N peaks at around 63 ($\log(S/N)\sim1.8$), but has a median value of 37 due to a significant low-S/N tail. The isodensity contours encompass 68.3\%, 95\%, and 99.7\% of the sample stars, respectively, and the colors delineate the relative stellar number density, with darker color representing higher density. }
\label{fig:fig2}%
\end{figure}

We define the hot star candidates by the following criteria,
\begin{itemize}
 \item[i] We adopt in LAMOST DR6 stars that are classified as O, B, and A types  according to the LAMOST pipeline. This contains 575,736 spectra of 408,544 unique stars. \\
 \item[ii] We cross-match the LAMOST DR6 with a full-sky hot star catalog (candidate OBA stars), devised by the same approach as \citet{Zari2021}, which is based on parallax, magnitudes, and colors from Gaia DR2 \citep{Brown2018} and 2MASS \citep{Skrutskie2006}. The key difference is that \citet{Zari2021} based their OBA star catalog on Gaia eDR3 \citep{Brown2021}. For our study, we have adopted Gaia DR2 (consistent with LAMOST DR6), and have selected hot stars with a 2MASS K$_s$-band absolute magnitude $M_{\rm K}<3$\,mag. The cross-match leads to 866,780 spectra of 590,341 unique stars in common. \\
 \item[iii] We adopt the LAMOST DR5 OB star catalog by \citet{LiuZ2019}. This contains 26,000 spectra of 16,002 unique stars.     
\end{itemize}
A simple union of these three source lists lead to 1,163,410 spectra of 844,790 hot star candidates.  Fig.\,\ref{fig:fig1} shows the distribution of these stars in Galactic coordinates, along with the LAMOST survey footprint. Unsurprisingly, these hot star candidates are concentrated to the Galactic plane, which is expected since most hot stars are young stars. Many of the stars at high Galactic latitudes turn out to be hot old stars, such as hot subdwarfs (sdBs, sdOs) and blue horizontal branch (BHB) stars. The majority of the stars are towards the Galactic anti-center, as targeted by the LSS-GAC. There are also a moderate number of stars in the first and second Galactic quadrants towards the inner Galaxy, particularly in the $Kepler$ field. Fig.\,\ref{fig:fig2} shows the distribution of the Gaia $G$-band magnitude and the LAMOST spectral signal-to-noise ratio (S/N) for our sample of candidate stars. The $G-$band magnitudes cover a dynamic range of nearly 15\,mag, and most of them are between $\sim$10 and 18\,mag, with a peak at around 14\,mag. The S/N peaks at about 63 ($\log(S/N)\sim1.8$), but the median value is about 37 due to a significant low-S/N tail.       

We correct for the line-of-sight velocity derived from the LAMOST pipeline \citep[e.g.][]{Luo2015}. All model fitting in this study will act on the normalized spectra. Specifically, all spectra were normalized by a smoothed version of themselves, derived by convolving the spectra with a Gaussian kernel of 50{\AA} width:
\begin{equation}
\bar{f}(\lambda_i) = \frac{\Sigma_jf_j\omega_j(\lambda_i)}{\Sigma_j\omega_j(\lambda_i)},
\end{equation}
\begin{equation}
\omega_j(\lambda_i) = \exp\left(-\frac{(\lambda_j-\lambda_i)^2}{L^2}\right),
L = 50\AA.
\end{equation}

\noindent
In this process we masked the Hydrogen lines, some other strong lines and the bands of interstellar absorption, by setting the weight of those pixels to be zero. 

\section{The stellar label information content of low-resolution, hot-star spectra} \label{sec:crbound}

Understanding what parameters and abundances we should adopt for our spectral grid requires us to evaluate the information content of these spectra: i.e., for which labels do the spectra vary appreciably when labels vary? And how covariate they are? One way to answer this question is via calculating the Cram\'er-Rao (CR) bound \citep{Cramer1945, Rao1945}. The same approached was demonstrated by \citet{Ting2016, Ting+2017a} and \citet{Sandford2020}, who applied the Cram\'er-Rao bound to predict the (theoretical) precision of stellar label determination of FGK stars for different spectral resolutions and wavelength coverage. 

The theoretical precision limit for a label estimate from a given spectrum depends on two aspects: (a) the \emph{gradient function} of the spectra $\nabla_l:=$ $\partial{f_\lambda}$/$\partial${\bf $l$}, which describes the response of the spectral flux $f_\lambda$ with respect to the change of stellar label {\bf $l$}, and (b) the flux uncertainties and covariances of the spectra. Following \citet{Ting+2017a}, the Cram\'er-Rao bound thus can be described as  
\begin{equation}
    K_{ij}^{-1} = \nabla_{i}f(\lambda_1)C^{-1}_{\lambda_1,\lambda_2}\nabla_{j}f(\lambda_2),
    \label{eq:CRbound}
\end{equation}
where $K_{ij}^{-1}$ is the inverse of the covariance matrix of the stellar labels, $C^{-1}$ the inverse of the covariance of the normalized spectral flux, $\nabla_if(\lambda_1)$ the gradient spectra with respect to label $i$ at $\lambda_1$, and  $\nabla_jf(\lambda_2)$ the gradient spectra with respect to label $j$ at $\lambda_2$. The gradient spectra are calculated through finite model differencing.

\begin{figure*}[hbt!]
\centering
\includegraphics[width=1.0\textwidth]{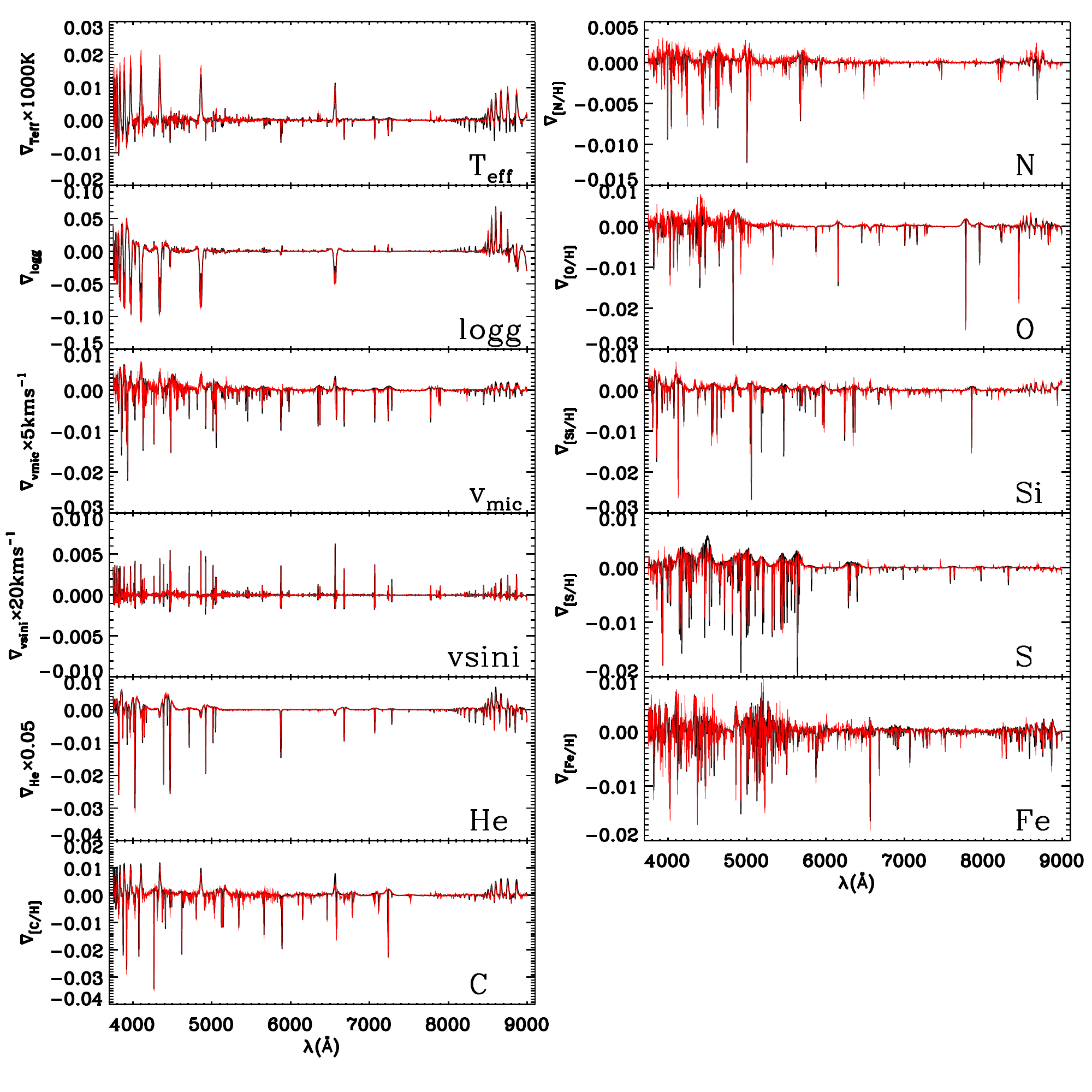}
\caption{The information content of low-resolution spectra for a B star. Each panel shows the gradient spectrum $\nabla_l:=\partial$$f_\lambda$/$\partial$$l$ with respect to different labels $l$, derived from finite differencing of Kurucz model spectra (black) at $\teff=18,000$\,K, $\logg=4.0$, $\feh=0$, and $\vsini=100$\,km/s. The gradient spectra generated from the \hotpayne are overplotted in red, illustrating the robustness of \hotpayne: the good agreement to the Kurucz model suggest that the \hotpayne determines the labels from the relevant features, instead of harnessing the astrophysical correlations. For \teff, \vmic, \vsini, and \he, the gradient spectra are multiplied by 1000\,K in \teff, 5km/s in \vmic, 20km/s in \vsini, and 0.05 in \he for better visibility. This Figure shows that the diagnostic signatures of label changes are widely distributed across the optical spectral range. Nonetheless, most spectral features are subtle due to the LAMOST's spectral resolution and the high effective temperature of the star: most abundance signatures are at the 1\%-level, even for abundance changes of 1~dex. Useful information about hot stars can only be extracted from a global fit to the spectrum and a careful spectral modeling.}
\label{fig:fig3}
\end{figure*}

As an example, Fig.\,\ref{fig:fig3} shows the gradient spectra $\partial$$f_\lambda$/$\partial$$l$ computed for a fiducial B star with $\teff=18,000$~K, $\logg=4.0$, and $\feh=0$. The figure illustrates that hot stars are information rich. Besides the stellar parameters, there are numerous pixels in the LAMOST wavelength range (3800--9000\,{\AA}) that are informative for a few selected elements. Unsurprisingly,  the most informative pixels for \teff and \logg are the Hydrogen lines, both Balmer and Paschen series, with moderate contributions from Helium lines and metal lines, whereas the information for \vsini resides mostly at the cores of strong absorption lines. . 
We also note that, the spectral features do not necessary need to come from the atomic transition of a species. For some elements, such as the He abundance, the change of which can modify the overall atmosphere structure appreciably, which in turn, indirectly modify the strength of spectral features. 

Although the figure demonstrates that many pixels contain valuable information for various labels, nonetheless the absolute changes for individual pixels are small in most cases: $|\partial\mathrm{f}_\lambda / \partial\mathrm{l}|$ changes rarely more than 1\% even for abundance changes of a full {\it dex}. These subtle and mostly covariate changes of spectral features are what inspire us to adopt full spectral fitting technique such as the \payne in the first place (see Section\,\ref{hotpayne}).

A realistic calculation of CR bound requires not only gradient spectra but also the assumed uncertainties of the spectral fluxes. For simplicity, for this theoretical exercise, we assume that the spectral measurements are independent among the adjacent pixels. We adopt the S/N (which vary as a function of the wavelength) of a typical LAMOST B star and scale the value such that we have 4650\,{\AA} to S/N = 100. Some assumptions of the CR bound calculation is undoubtedly optimistic. For example, the uncertainties for the adjacent pixels of an observed spectrum can be correlated. Furthermore, the calculation assumes a perfect knowledge of the spectral models. But we emphasize that this exercise aims only to determine what labels we should adopt for the spectral grid. On top of that, it has been shown that the CR bound calculations can usually be attained within a factor of two despite these imperfections \citep[e.g.][]{Ting+2019, Xiang2019}. 

\begin{figure*}[hbt!]
\centering
\includegraphics[width=1.0\textwidth]{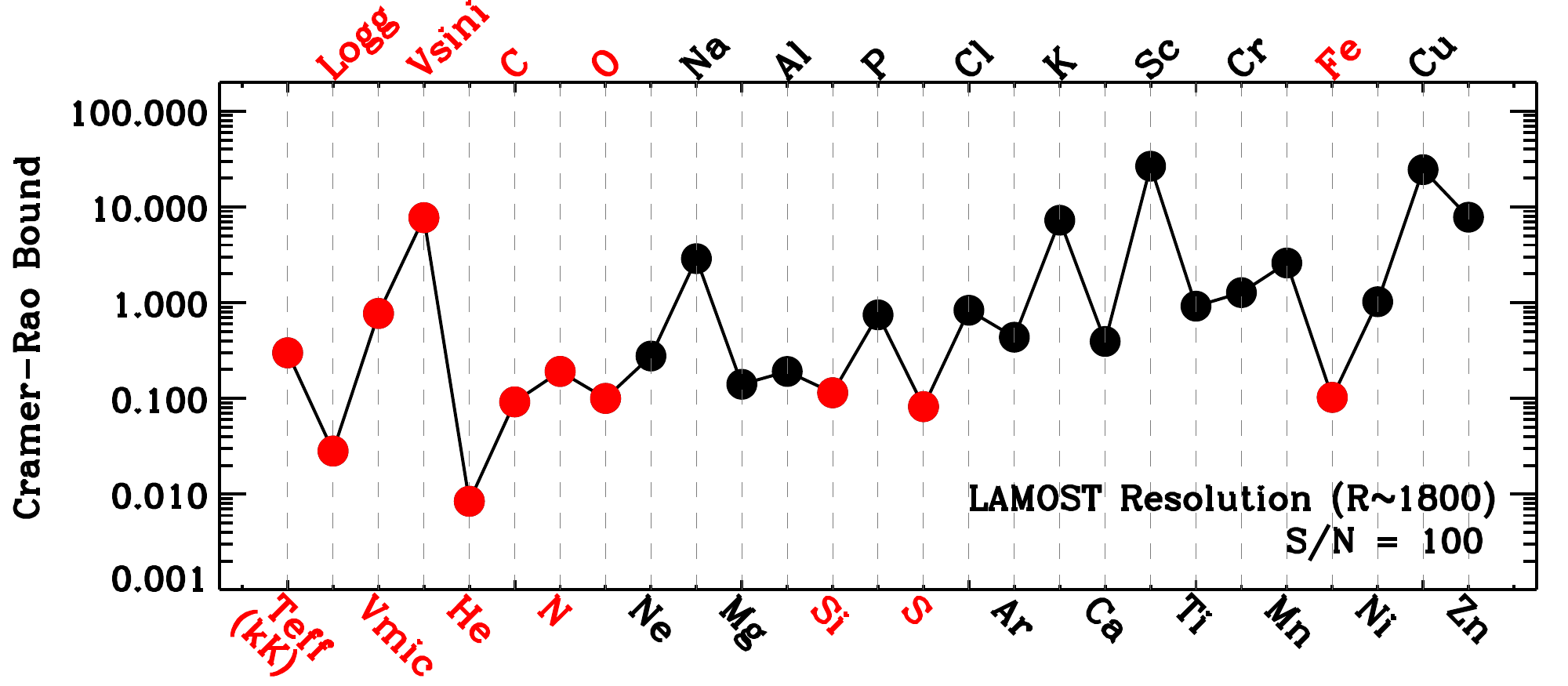}
\caption{Theoretical precision limits for stellar label determination from LAMOST of hot stars, expressed through the Cram\'er-Rao (CR) bound, $\sqrt{K_{ii}}$ (Eq.~\ref{eq:CRbound}), for a fiducial B star with $\teff=18000$\,K, $\logg=4.0$, and $\feh=0$, \vsini$=$100\,km/s. The LSF and S/N of the Kurucz model spectra as a function of wavelength are matched to that of a LAMOST B star spectrum, with S/N = 100 at 4650{\AA}. The CR bound is expected to vary with 1/(S/N). The units of the CR bound on the Y-axis differ among different labels: they are 1000\,K for \teff, cm/s$^{2}$ for \logg, km/s for $v_{\rm mic}$ and $v{\rm sin}i$, {\rm dex} for abundance [X/H] except for He. For He, the number indicates the atomic number ratio \he. Labels shown in red are those we attempt to derive from the LAMOST spectra in this work.}
\label{fig:fig4}%
\end{figure*}

The CR bound, expressed by the diagonal terms of the Cram\'er-Rao matrix $\sqrt{K_{ii}}$, is shown in Fig.\,\ref{fig:fig4} for our fiducial B star spectrum. The calculation predicts that, in the limit of spectral perfect knowledge, we should obtain measurements with precision of $\sim$300\,K for \teff, 0.03\,dex for \logg, 1\% for $N_{\rm He}/N_{\rm tot}$, 1\,km/s for $v_{\rm mic}$, and 8\,km/s for $v{\rm sin}i$ from a typical LAMOST hot-star spectrum. The CR bound for abundances is within 0.1\,dex for a few elements with strong features, such as C, O, Si, and Fe. For other elements such as N, Mg, and Al, the CR bounds are within 0.2\,dex. Note that although the CR bound for only some selected elements (with atomic number no larger than Zn) are presented in the Figure (we omit all CR bounds that are larger than 10\,dex), all the elements with atomic number smaller than Eu, except for Li, Be, B, F, V, and Co, are included for the CR bound computation. The Li, Be, B, F, V, and Co are discarded because their null gradients could cause numerical artefacts in the results. Finally, we note that here we assume the B star as a representative reference point for this study, but clearly, the CR bound will vary for different stellar types.

\begin{figure*}[hbt!]
\centering
\includegraphics[width=0.8\textwidth]{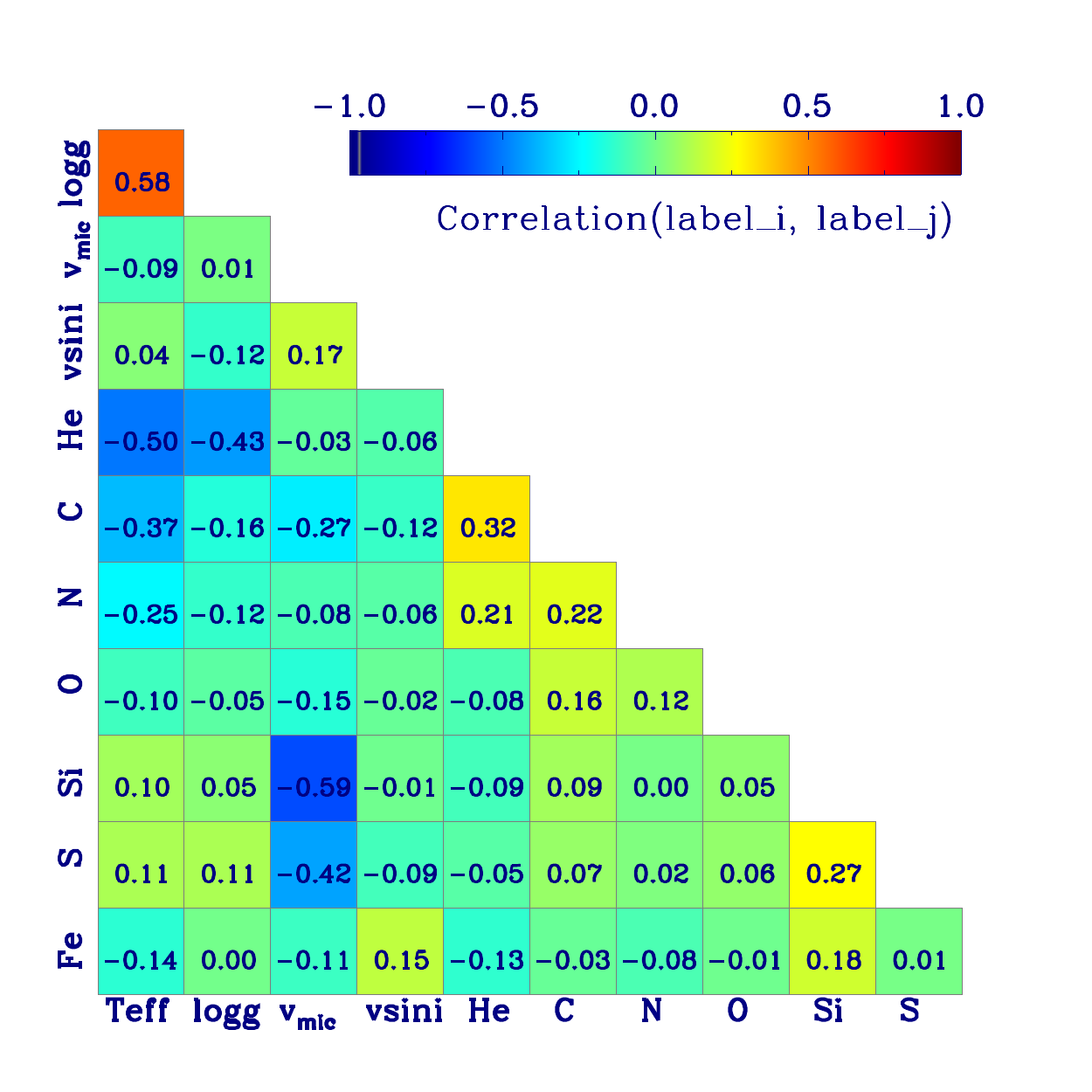}
\caption{Correlation coefficients among the stellar label estimates, predicted by the Cram\'er-Rao bound calculation (see Eq.~\ref{eq:CRbound}). These correlations were derived for the spectrum of a fiducial B-type star with $\teff=18000$\,K, $\logg=4.0$, $\feh=0$, and $\vsini=100$\,km/s. Prominent among the correlations are those between estimates of \teff and of \logg, and between $N_{\rm He}/N_{\rm tot}$ and both \teff and \logg. The nontrivial correlations suggest that the most  spectral features are somewhat blended/degenerate and thus call for a \emph{simultaneous} fit of all pertinent labels to the entire spectrum.}
\label{fig:fig5}%
\end{figure*}
Fig.\,\ref{fig:fig5} presents the covariances among the selected labels, reflecting in the off-diagonal term of the Cram\'er-Rao matrix $K_{ij}$. As expected, the \teff and \logg are strongly correlated, as their gradient spectra share many common spectral features (H and He lines). Similarly, the $N_{\rm He}/N_{\rm tot}$ is strongly \emph{anti-}correlated with \teff and \logg. The [Si/H] and [S/H] are strongly correlated with $v_{\rm mic}$. The [C/H] and [N/H] are moderately correlated with \teff. The $v{\rm sin}i$, [O/H] and [Fe/H] are not strongly correlated with any of the other labels. The non-trivial correlation between labels due to somewhat degenerate spectral features further demonstrates that the pertinent labels can be only be obtained by fitting all labels \emph{simultaneously} through full spectral fitting.

To sum up, inspired by this exploration of the CR bounds for low-resolution spectra of hot stars, we choose to build our spectral model for these LAMOST spectra that depends on 11 labels, marked in red in Fig.\,\ref{fig:fig4}, namely \teff, \logg, $v_{\rm mic}$, $v{\rm sin}i$, $N_{\rm He}/N_{\rm tot}$, [C/H], [N/H], [O/H], [Si/H], [S/H], and [Fe/H] (Section\,4), as they are among labels we may derive from the spectra with the best precision.

\section{Stellar label determination}\label{sec:getting_stellar_labels}
In this section, we introduce our method of determining stellar labels for hot stars from the LAMOST spectra. This includes an introduction of the Kurucz model spectral library, the application of the \hotpayne for the spectral fitting, as well as examinations of the resultant label estimates. 

\subsection{The Kurucz synthetic spectral library}
\begin{table}
\caption{Range and sampling method of the stellar labels of Kurucz ATLAS12 models, calculated for the \hotpayne spectral modelling (see Section~\ref{sec:getting_stellar_labels}). The MIST sampling method means we draw from the MIST grid with equal EEP.}
\label{table:table1}
\begin{tabular}{lll}
\hline
 Labels &    range  &  sampling  \\
\hline
 \teff &  [7500, 60000] (K) & MIST \\
 \logg & [0, 5]  & MIST \\
 \feh & [$-3$,0.5] & MIST \\
 $v_{mic}$ & [0, 15] (km/s) & uniform \\
 \vsini & [0, 500] (km/s) & uniform \\
 $N_{\rm He}/N_{\rm total}$ & [0, 0.25] & uniform \\
 ${\rm [C/Fe]}$ & [$-2$,3] & uniform \\
 ${\rm [N/Fe]}$ & [$-2$,3] & uniform \\
 ${\rm [O/Fe]}$ & [$-2$,3] & uniform \\
 ${\rm [Si/Fe]}$ & [$-1$,2] & uniform \\
 ${\rm [S/Fe]}$ & [$-1$,2] & uniform \\
 \hline
\end{tabular}
\end{table} 
We build a library of synthetic spectra generated with {\sc SYNTHE}, using the Kurucz LTE atmospheric models calculated with ATLAS12 \citep{Kurucz1970, Kurucz1993, Kurucz2005}. We consider 11 physical labels, \teff, \logg, and \feh, $v_{mic}$, $v{\rm sin}i$, $N_{\rm He}/N_{\rm tot}$, [X/H] for X=C, N, O, Si, S for the spectral modelling. The parameter ranges are presented in Table\,\ref{table:table1}. For \teff, \logg, and \feh, we sample the grids from the MIST isochrones \citep{Paxton2011, Choi2016} uniformly in equivalent evolutionary phase (EEP). While the other labels are uniformly drawn within the given range for each label. For all other abundances, we simply scale them with \feh, assuming the solar abundance scale of \citet{Asplund2009}. 
The model spectra are broadened according to the \vsini value. To ensure the convergence of the model computation, we require that at both Rosseland mean opacities $\tau_{\rm ross}=0.1$ and $\tau_{\rm ross}=1$ the flux errors to be smaller than 2\%, and the flux derivative errors to be smaller than 20\%. Also, we require the maximum variations of temperature for the inner atmospheric layers (40-80) between two iterations to be fractionally smaller than $10^{-4}$.

In total, we generate 10,127 Kurucz model spectra. The native model spectra generated with the SYNTHE have a resolving power of $R=500,000$, and we degrade them to match the line spread function (LSF) of the LAMOST spectra. We adopt a Gaussian profile for the LSF, the width of which is a wavelength-dependent function, approximated by a 2-order polynomial for the blue- and red-arm, separately. The LAMOST spectra have $R\sim1800$ at 5000\,{\AA} (FWHM$\sim$2.8\,{\AA}), but the detailed spectral LSF is found to vary fiber to fiber and exposure to exposure \citep[e.g.][]{Xiang2015}. 
To generate a range of realistic LSFs for the model grid, we assign a specific LSF for each model spectrum, using individual LAMOST LSFs derived from different exposures and fibers.
The spectra are further broadened to their corresponding \vsini assuming rotational profiles with a constant limb darkening coefficient of 0.6 \citep{Gray1992,Hubeny2011}. 

Our LTE Kurucz models come with two caveats:
\begin{itemize}
  \item For very hot ($\gtrsim$40,000\,K) stars with either relatively high metallicity ($\feh\gtrsim-0.25$\,dex) or relatively low \logg values ($\lesssim$4.0), those with very high CNO abundance values can lead to a low Hydrogen fraction (e.g. $<0.6$). In this case, the model computations are hardly converged, which leads to a poor coverage in those parts of the parameter space.
  \item The model spectra are generated using the Kurucz 1D LTE atmospheric models. As metioned in Section\,1, it is known that the NLTE effect and stellar winds are non-negligible for hot stars, and especially strong for giants \citep[e.g.][]{Kurucz1970, Kudritzki1979, Kudritzki1998, Hubeny1995, Lanz2007, Hainich2019}. In this work, we chose to apply the LTE model as a proof of concept. While the Kurucz  model spectra may be less accurate than NLTE spectra, it is flexible to obtain self-consistently computed model spectra within a broad parameter ranges as desired and with relatively complete line lists. Nonetheless, as we will demonstrate in the rest  of the paper, further efforts based on NLTE model spectra are necessary in order to generate accurate elemental abundances. 
\end{itemize}

\subsection{The \hotpayne} \label{hotpayne}
The \payne is a full spectral fitting tool for stellar label determination built on several key ingredients: First, it trains a flexible spectral interpolation model based on neural network that allows one to interpolate spectra precisely in high dimensions (e.g. $\sim20$), Furthermore, The \payne emphasizes the importance to generate models $self-consistent$ly. The self-consistency here refers to the fact that, for any changes in stellar labels (including all the stellar labels), we always solve for the atmospheric structure before running the radiative transport. Secondly, the \payne fits all labels of consideration simultaneously from the full spectra with the goal to make full use of the information in the spectra and to properly characterize the covariances of labels arose due to line blending effects. Such effect is particularly prominent for low-resolution spectroscopy \citep{Ting+2019}.

For training the \payne, we divide the Kurucz spectral library into two sets, with 9115 spectra as the `training set', and the other set with 1012 spectra as the validation test set. We train the neural-network spectral interpolation model pixel-by-pixel. That is, for each pixel, we train a neural network for the flux as a function of the labels considered. Our neural network model has two layers, each with 100 neurons, and the Sigmoid activation function is adopted to scale the input spectra. The training is carried out in the environment of $Pytorch$, using the adaptive moment (ADAM) estimation algorithm \citep{Kingma2014}. 
In the training process, we have 17 free parameters, including the 11 target stellar labels shown in Table\,\ref{table:table1}, and six more coefficients describing the LSF (three for each of the blue and red arm, respectively) as introduced above. We also explored the possibility of deriving macroturbulent velocity $v_{\rm macro}$, which reflects the contribution of non-rotational broadening to the line profiles \citep[e.g.][]{Simon-Diaz2014}, on top of \vsini, but we found that the two parameters are too degenerate at the LAMOST resolution. We chose to only fit for \vsini in this work. Consequently, our \vsini estimates are likely contributed by both the rotational broadening and macroturbulence. Our estimates might overestimate the true \vsini in case where the non-rotational broadening is comparable or stronger than the rotational broadening. For example, \citet{Simon-Diaz2014} found that for stars with $\vsini\lesssim120$\,km/s, ignoring $v_{\rm macro}$ may lead to an overestimate of \vsini by $\sim$25\,km/s in general.

\begin{figure*}[hbt!]
\centering
\includegraphics[width=1.0\textwidth]{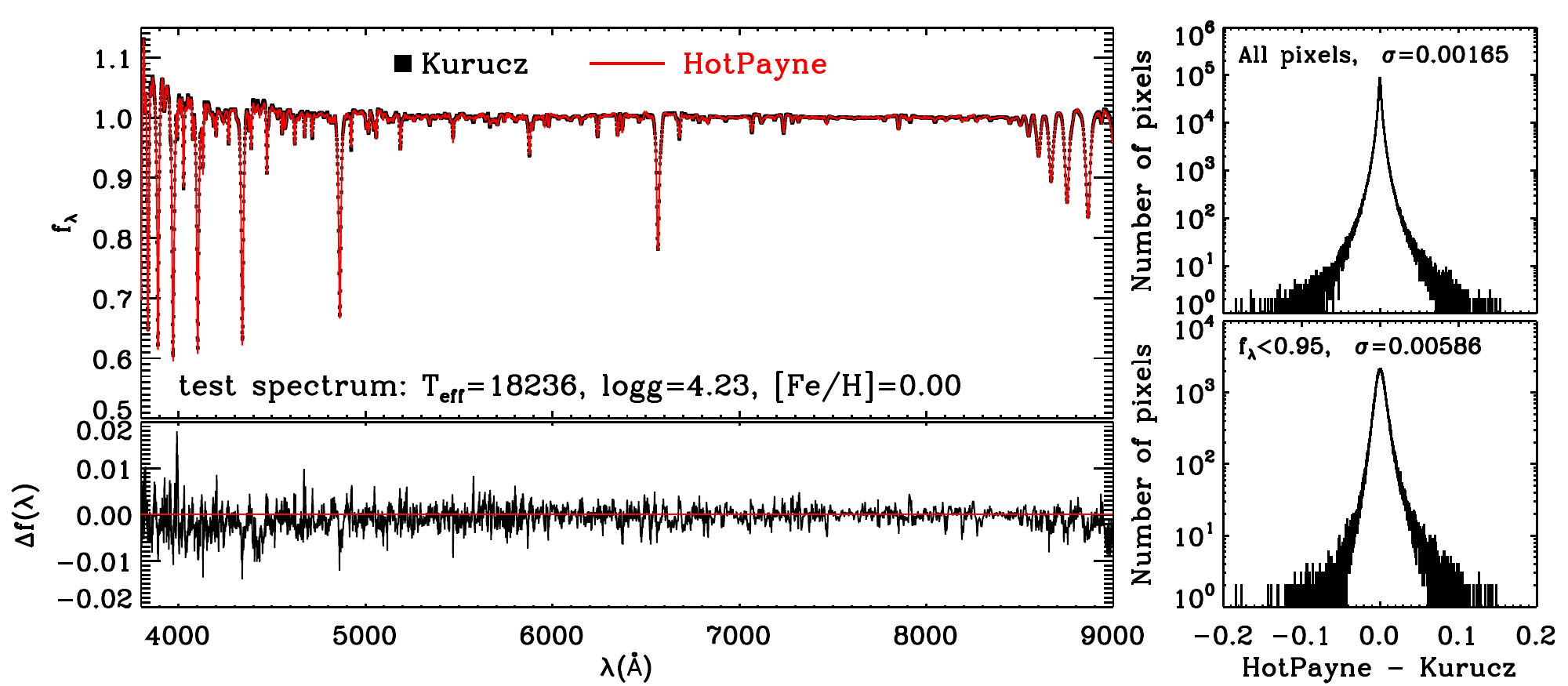}
\caption{Precision of the spectrum prediction by the \hotpayne, which entails a neural-net enabled interpolation among Kurucz model spectra. The left panels show the comparison between the Kurucz spectrum and its prediction by the \hotpayne for a test star with \teff=18236\,K, \logg=4.23, and \feh=0.0. The top-right panel shows the distribution of flux residuals for all of our 1012 test spectra, each with 3801 pixels in the wavelength range of
$\lambda\lambda$3750--9000{\AA}. A Gaussian fit to the distribution yields a standard deviation of 0.165\%. The bottom-right panel is the analogous to the top-right panel, but only for pixels whose normalized spectral flux are
smaller than 0.95, i.e., for strong absorption lines, mostly the cores of H and He lines. The standard deviation from a Gaussian fit is 0.586\% for the strong absorption features.}
\label{fig:fig6}%
\end{figure*}
Fig.\,\ref{fig:fig6} illustrates the precision of the spectral interpolation by the \hotpayne deduced using the test spectra. For all individual pixels of the 1012 test spectra, the overall dispersion of the residuals between the \hotpayne fits and the Kurucz model flux is 0.165\%, which is equivalent to a spectral S/N of 600. For those pixels with normalized flux values smaller than 0.95, mostly the cores of strong H and He lines, the dispersion of the residuals is 0.586\% (or equivalently, S/N of 200). In other words, the interpolation errors from the \payne is largely negligible given the typical S/N of LAMOST is lower than 200. 

\begin{figure*}[hbt!]
 \centering
 \includegraphics[width=0.9\textwidth]{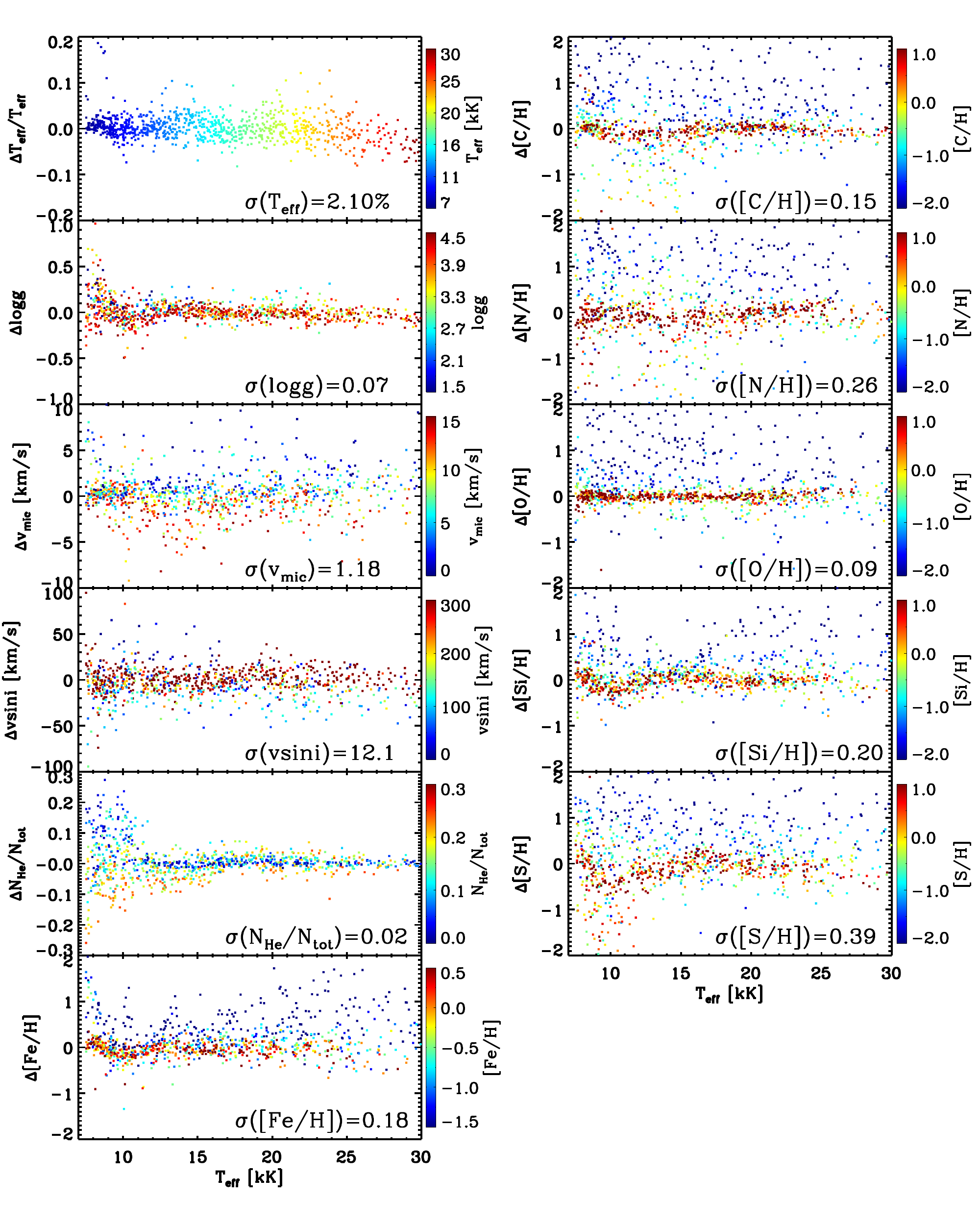}
 \caption{Precision of the \hotpayne's label estimates as a function of $\teff$. Plotted are the differences (residuals) between the labels from the \hotpayne and the true values for a test set of Kurucz model spectra with $S/N\sim100$, color coded by the true values. The numbers marked in the figure are the dispersions derived from a Gaussian fit to the residuals. The precision of the \hotpayne's label determination varies with \teff and [X/H]. For metal-poor stars of ${\rm [X/H]}<-1$\,dex, the [X/H] estimates may suffer large uncertainties. However, we note that in the real case of LAMOST survey, we do not expect there are many hot stars with such low metallicity.} 
 \label{fig:fig7}%
\end{figure*}
To assess the internal precision of stellar label determination with the \hotpayne, we make a sanity check using the Kurucz test spectral sample. That is, we derive the stellar labels of the Kurucz test spectra with the \hotpayne, and compare with the true labels. We add noise to the test spectra (with S/N$\sim$100 at 4650{\AA}), as for the CR bound calculations. For reason that we will elaborate in Section\,\ref{sec:lsf}, we adopt a fixed LSF by taking the mean LAMOST LSF, and do not fit for them, the same way we will analyze the LAMOST data.

Fig.\,\ref{fig:fig7} plots the differences (residuals) between the \hotpayne stellar label estimates and the true values as a function of the true \teff, and color coded by the true values of the labels. On average, the model-to-model fits show that we can recover the stellar labels, except for the very metal poor end. A Gaussian fit to the residuals suggest that, on the whole, the internal precision is about 2.1\% for \teff, 0.07\,dex for \logg, 1.2\,km/s for \vmic, 12.1\,km/s for \vsini, 0.02 for \he, 0.09\,dex for [O/H], about 0.15\,dex for [C/H], 0.18\,dex for [Fe/H], 0.20\,dex for [Si/H], 0.26\,dex for [N/H], and 0.39\,dex for [S/H]. We note that for some labels such as [N/H] and [S/H], the error is partially dominated by a few outliers, a consequence of the strong variations of label precision across the parameter (\teff and [X/H]) space. For stars with $\teff\sim18,000$\,K and ${\rm [Fe/H]}\sim0$\,dex, we find that the precision is consistent with the CR bound (Fig.\,\ref{fig:fig4}) within a factor of 2.

The precision of the \hotpayne's label estimates vary significantly with both \teff and [X/H]. For stars with $\teff<11,000$\,K, the \logg residuals exhibit a strong negative trend with \teff, causing a \teff-dependent systematic bias in \logg of up to about $0.2$\,dex at the low-\teff end. Similar trend is also presented in the \feh, [Si/H], and [S/H] panels, causing a \teff-dependent systematics of about $0.1$\,dex in \feh, and larger values in [Si/H] and [S/H]. This trend reflects an imperfection of our spectral emulation for this cool part of the parameter space, probably due to either insufficiently flexible neural networks or non-optimal training process (e.g. gets stuck in a local minimum). There are also many more metal lines in the low-temperature spectra, which increase significantly the complexity of the emulation. At this cool end, our test shows that the \he estimates can be challenging but the \he estimates are robust for stars with $\teff>11,000$\,K, because the He lines are more prominent at this temperature range. 
For the metal-poor stars of ${\rm [X/H]}\lesssim-1$\,dex, the abundance estimates may exhibit large deviations from the true values, especially for [C/H], [N/H], [O/H], and [S/H]. This is just the manifestation that for metal-poor stars, the spectral features become weak to allow for any abundance estimates. However, we note that in the real case of LAMOST survey, we do not expect there are many hot stars with such low metallicity, except for some chemical peculiars with very low CNO abundances due to atomic diffusion \citep[e.g.][]{Michaud2015}.

While the model-to-model fit remains optimistic, we caution that when fit to the LAMOST spectra, the fit is likely to incur larger uncertainties due to the mismatch between model and observed spectra due to model imperfections.

\subsection{The line spread function} \label{sec:lsf}
The LSF of the individual LAMOST spectra vary with fibers and exposures, which is expect to have an impact on the stellar label determination \citep[e.g.][]{Xiang2015}. In design, our \hotpayne module can fit the LSF from the spectra simultaneously with the stellar labels. However, we found that to fit the LSF from the LAMOST spectra is challenging as it is easily impacted by any small mismatch of the line cores between the observed LAMOST spectra and the Kurucz model spectra, due to systematics of the latter. A more sensible way is to take the LSF of each LAMOST spectrum  derived from arc lamps and sky emission lines as input, and fix it in the \hotpayne fitting. However, currently such LSF data for LAMOST spectra are still missing.  

As a compromise, we instead adopt the mean LSF of the LAMOST spectra derived by the author using arc lamps and sky emission lines from the LSS-GAC DR2 \citep{Xiang2017b}. We test the impact of the fiber-to-fiber and exposure-to-exposure variations LSF on the stellar label determination by assigning each Kurucz test spectrum (Section\,\ref{hotpayne}) with a unique LSF from the LSS-GAC DR2, and fit them with the \hotpayne model spectra with the mean LSF. We find that ignoring the of variations of LSF will cause larger uncertainties on the \vsini estimates, which is 17\,km/s (compared to the 12\,km/s from the mean LSF as shown in Fig.\,\ref{fig:fig7}), as a consequence of covariance between the LSF and \vsini. For a minor number of spectra with very different LSF, the \vsini estimates may differ by larger than 30\,km/s from the results of the mean LSF. Whereas for the other labels, the LSF has only marginally effects on the \hotpayne's determination in most cases.

\subsection{Initial estimates of \teff from line indices}
To facilitate the spectral fitting with the \hotpayne to LAMOST data set, we first make an initial estimate of \teff based on the indices (equivalent widths) of the Ca\,{\sc ii} 3933\,{\AA}, H\,{\sc i}+Ca\,{\sc ii} 3967\,{\AA}, and H\,{\sc i} 4101\,{\AA} lines. To compute the line indices, we adopt the $\lambda$3912--3922\,{\AA} and $\lambda$3946--3956\,{\AA} windows for continuum estimates of the Ca\,{\sc ii} 3933 line. Similarly, the continuum for H\,{\sc i}+Ca\,{\sc ii} 3967\,{\AA} is derived from the $\lambda$3946--3956\,{\AA} and $\lambda$3980--3990\,{\AA} windows, and the continuum for H\,{\sc i} 4101\,{\AA} line is derived from the $\lambda$4080-4090\,{\AA} and $\lambda$4114--4124\,{\AA} windows. 

We use a subset of 57,039 LAMOST spectra as the reference set to build an empirical relation between line indices and \teff. This reference set includes 16,002 stars from the LAMOST OB sample of \citet{LiuZ2019}, and the remaining ones are selected from the LAMOST and \citet{Zari2021} common star sample, the majority of which are A- and F-type stars according to the LAMOST classification. The \teff of these reference spectra are estimated with \hotpayne assuming a initial \teff set to be the mean \teff of our training library spectra. For late F-type stars, the \teff estimates with the \hotpayne might be inaccurate asthey are outside the \teff coverage of the Kurucz training set, i.e., cooler than $<7500$\,K. We therefore adopt the \teff provided in \citet{Xiang2019} if it has smaller $\chi^2$ than the \hotpayne fits. Fitting a 3-order polynomial function with this reference set to predict the \teff with line indices yields the following relation:
\begin{equation}
\begin{aligned}
\log{T_{\rm eff}} & =  4.71047 - 4.43455\times10^{-2}C_{3933} - 5.37380\times10^{-2}C_{3967} \\
& + 3.46889\times10^{-2}C_{4101} - 5.68345\times10^{-3}C_{3933}^2  \\
& - 1.26749\times10^{-2}C_{3967}^2 - 2.40555\times10^{-2}C_{4101}^2 \\
& + 1.26779\times10^{-2}C_{3933}C_{3967} - 4.19722\times10^{-2}C_{3933}C_{4101} \\
& + 2.34087\times10^{-2}C_{3967}C_{4101} \\
& + 8.44885\times10^{-4}C_{3933}C_{3967}C_{4101} \\
& - 9.98430\times10^{-4}C_{3933}^2C_{3967} + 2.23513\times10^{-3}C_{3933}^2C_{4101}  \\
& + 6.24433\times10^{-4}C_{3933}C_{3967}^2 + 1.40913\times10^{-3}C_{3933}C_{4101}^2 \\
& - 1.78560\times10^{-3}C_{3967}^2C_{4101} + 1.85542\times10^{-3}C_{3967}C_{4101}^2  \\
& + 5.01277\times10^{-4}C_{3933}^3 + 5.10836\times10^{-4}C_{3967}^2 \\
& + 1.52060\times10^{-4}C_{4101}^3,
\end{aligned}
\end{equation}
where $C_{3933}$, $C_{3967}$, and $C_{4101}$ are the equivalent widths of the Ca\,{\sc ii} 3933\,{\AA}, H\,{\sc i}+Ca\,{\sc ii} 3967\,{\AA}, and  H\,{\sc i} 4101\,{\AA} lines, respectively. For OB stars, the Ca line index may suffer from some uncertainties due to interstellar absorption, but since our estimates from the line indices only serve as an initial guess to optimize the running time, such uncertainties have minimal impact on our final estimates.

\subsection{Spectral masking and the \emph{spurious absence} of early A-type stars}
\begin{figure*}[hbt!]
\centering
\includegraphics[width=0.9\textwidth]{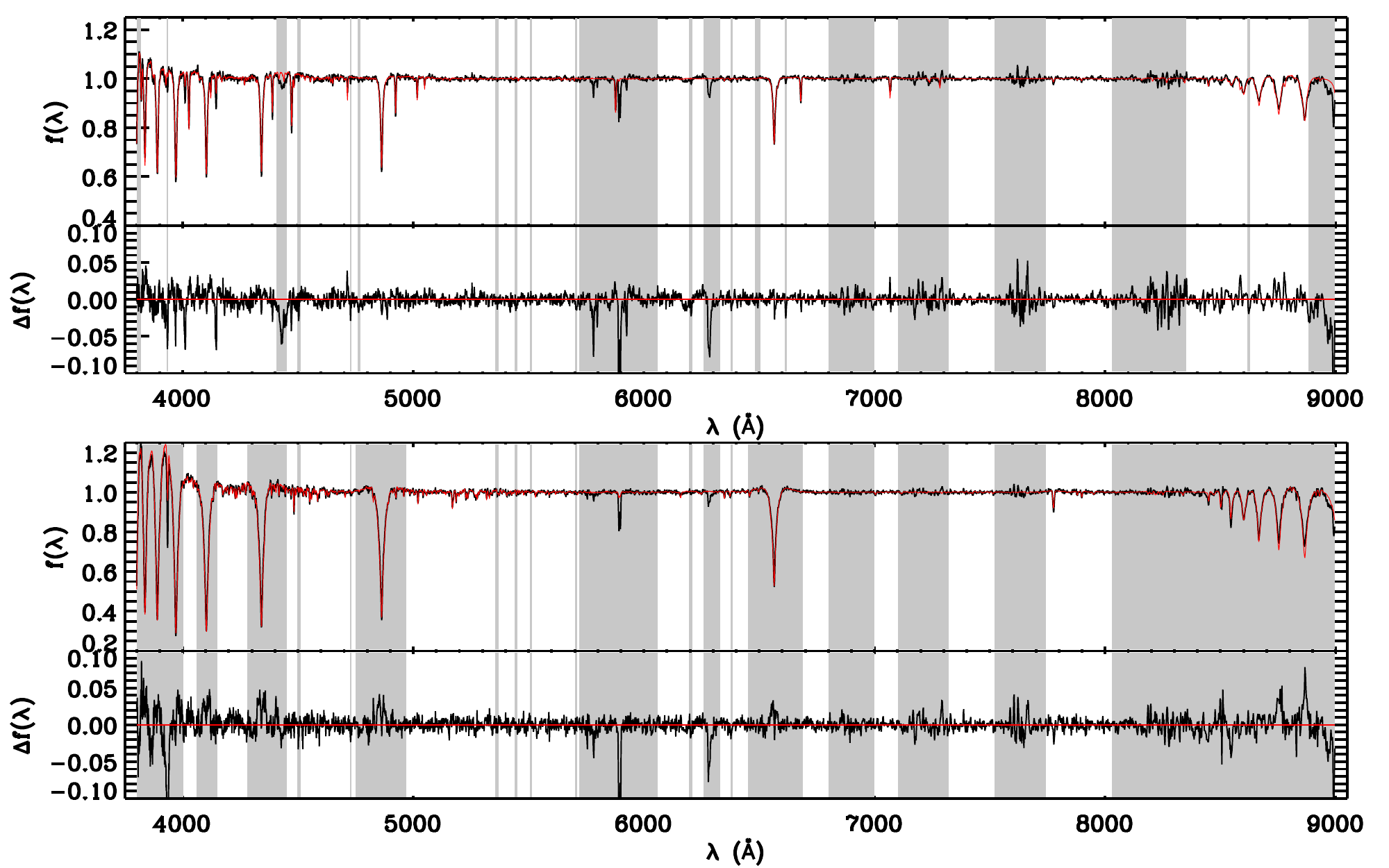}
\caption{An example of the \hotpayne fit for a B-type star (top) and an early A-type star (bottom). The LAMOST spectra are shown in black, and the best fit \hotpayne models are shown in red. The masked telluric bands, DIB, interstellar absorption of Na and Ca, and the dichroic region are marked by shaded regions.  For the case of early A-type stars, the wavelength windows of Hydrogen lines (Balmer \& Paschen series) are also masked (see text).}
\label{fig:fig8}%
\end{figure*}
When fitting the LAMOST spectra, we also mask the wavelength windows with known telluric or interstellar medium adsorptions, In particular,  we mask the known strong diffuse interstellar bands from the DIBs directory\footnote{http://dibdata.org} \cite[][]{Hobbs2008}, the Na\,{\sc i} $\lambda$5890,5896 doublet, and the Ca\,{\sc ii}\,K $\lambda$3933 line. We also discard the blue ($\lambda<3820$\,{\AA}) and red ($\lambda>8880$\,{\AA})  edges of the LAMOST spectra, as well as the dichroic region (5700--6050{\AA}) of the blue and red arms of the LAMOST spectrographs. The masked wavelength regions are shaded  in Fig.\,\ref{fig:fig8} . 

\begin{figure*}[hbt!]
 \centering
 \includegraphics[width=1.0\textwidth]{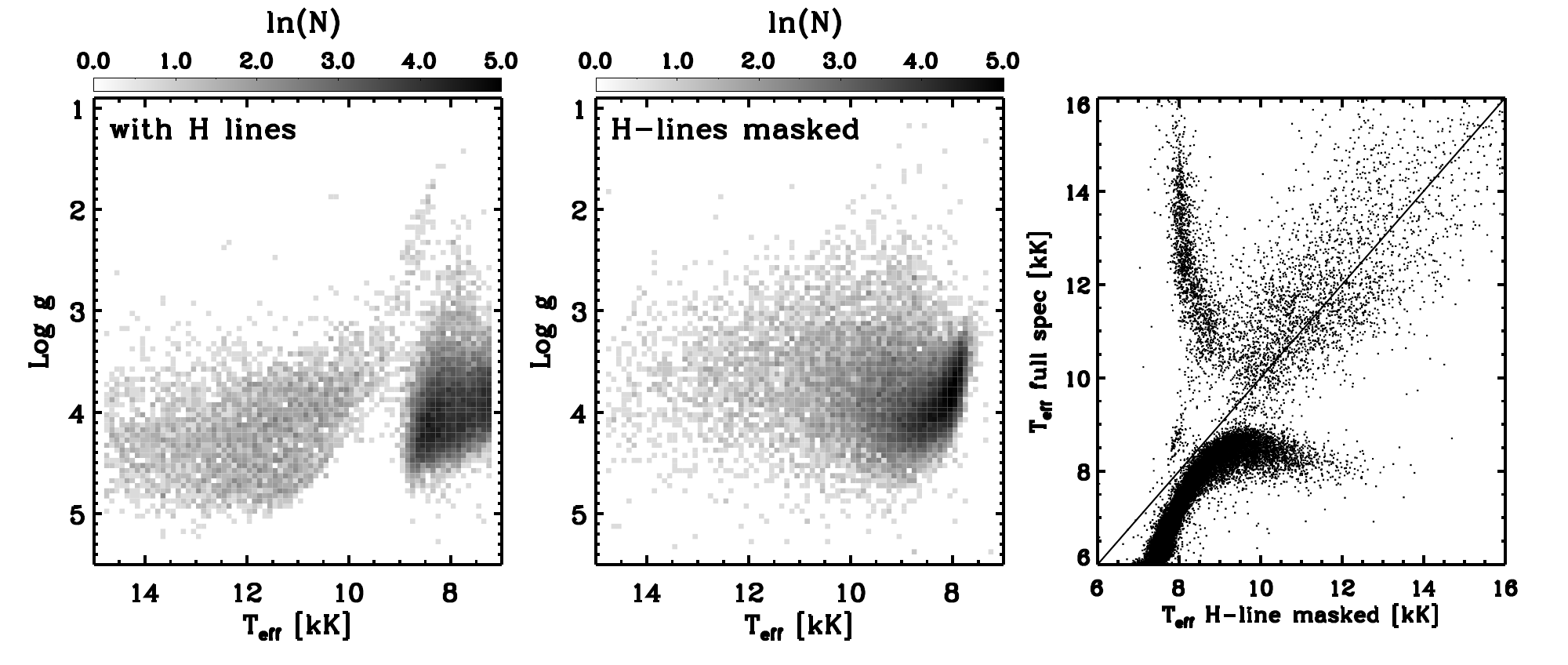}
 \caption{\teff-\logg diagrams for \teff derived from the LAMOST spectra without (left) and with (middle) masking the Hydrogen lines. The right panel shows a direct comparison of the \teff estimates in the left and in the middle panels. The inclusion of Hydrogen lines for the label determination causes an artificial desert of early A-type (and late B-type) dwarf stars in the \teff-\logg diagram, due to the reverse of spectral gradients $\partial{f}_\lambda/\partial\lambda$ for the A0 stars. Masking the Hydrogen lines resolves this issue (middle panel). In this work, we consider labels estimates by combining these two estimates, putting more weight on the masked estimates in the \teff range where the Hydrogen lines gradient reversal happens.} 
 \label{fig:fig9}%
\end{figure*}
Our preliminary fits lead to resultant stellar parameters that exhibit a desert of early A-type and late B-type dwarf stars ($9000\lesssim\teff\lesssim11,000$\,K) in the \teff--\logg diagram (left panel of Fig.\,\ref{fig:fig9}). This desert is unexpected in the context of Galactic evolution and star formation/evolution scenarios. A careful inspection on the spectral models reveals that this artefact is possibly related to the change of signs in the gradient spectra of Hydrogen lines for the A0 type stars. For A0 stars ($\teff\sim9500$\,K), the strength of Hydrogen lines reach the maximum, so that the gradient spectra of Hydrogen lines for A0 stars reach a minimal, with absolute values close to 0 (see Appendix). This causes an inflection point in the spectral gradient features of the Hydrogen lines as a function of \teff. As a consequence, the parameter estimates are sensitive to the initial \teff values. Depending on the initial guesses from the line indices, the fits tend to pile out due the gradient gap, causing the desert in the \teff-\logg diagram as seen in the left panel of Fig.\,\ref{fig:fig9}. These effects are illustrated in the right panel of Fig.\,\ref{fig:fig9}. 

As a remedy, we mask all the Hydrogen lines for label determination for these early A-type and late B-type stars. The masked regions are shown in the bottom panels of Fig.\,\ref{fig:fig8}. Masking the Hydrogen lines can reduce the spectral information content and lead to lower precision for the label determinations, particularly for \teff and \logg. Nonetheless, an examination of the CR bound for early A-type stars suggests that the Hydrogen-masked spectra are still able to yield robust estimates for many of the labels (see Appendix). The \teff--\logg diagram deduced for a subset of LAMOST stars after masking the Hydrogen lines is shown in the middle panel of Fig.\,\ref{fig:fig9}. Masking the Hydrogen lines resolved the non-sensible desert. In order to benefit from the full spectra fitting as we can while mitigating the impact of Hydrogen lines for the A0 stars, we combine the two sets of label estimates to give the ultimate label estimates\footnote{Since those with $\teff^{\rm H-masked}<9500$\,K \& $\teff^{\rm fullspec}>9500$\,K and $\teff^{\rm H-masked}>9500$\,K \& $\teff^{\rm fullspec}<9500$\,K are likely affected by imperfect initial \teff estimates from the line indices. For those objects, we opt to re-determine their full-spectra-based labels using the H-masked \teff as the initial estimates.}.  The combination is carried out with a weighted-mean method,
\begin{equation}
 \vec{l} = \frac{\omega_1\vec{l}_1 + \omega_2\vec{l}_2}{\omega_1 + \omega_2},
\end{equation}
where $\vec{l}_1$ and $\vec{l}_2$ represent the labels derived from spectra with and without the Hydrogen lines masked, respectively. We adopt the {\it Hann} function as the weights for labels derived from the H-masked spectra, 
\begin{equation}
 \omega_1(x) = 
  \begin{cases}
    \frac{1}{2}\left(1+\cos\left(\frac{2{\pi}x}{L}\right)\right) = \cos^2\left(\frac{{\pi}x}{L}\right), & \text{if } |x|\leq L/2 \\ 
    0, & \text{if } |x|>L/2,
  \end{cases}
\end{equation}
where $x = \teff - 9500$\,K, and the \teff is that from the H-masked spectra. The weights for the labels derived from the full spectra is
\begin{equation}
\omega_2 = 1 - \omega_1.
\end{equation}
As for the smoothing length, for stars with $\teff^{\rm fullspec}>9500$\,K, we adopt a smoothing length $L$ of 5000\,K, such that $\omega_1=0$ for stars with $\teff^{\rm H-masked}>12000$\,K. And for stars with $\teff^{\rm fullspec}<9500$\,K, we adopt a length $L$ of 3000\,K, such that $\omega_1=0$ for stars with $\teff^{\rm H-masked}<8000$\,K. 

\subsection{Examination with repeat observations}
\begin{figure*}[hbt!]
\centering
\includegraphics[width=0.8\textwidth]{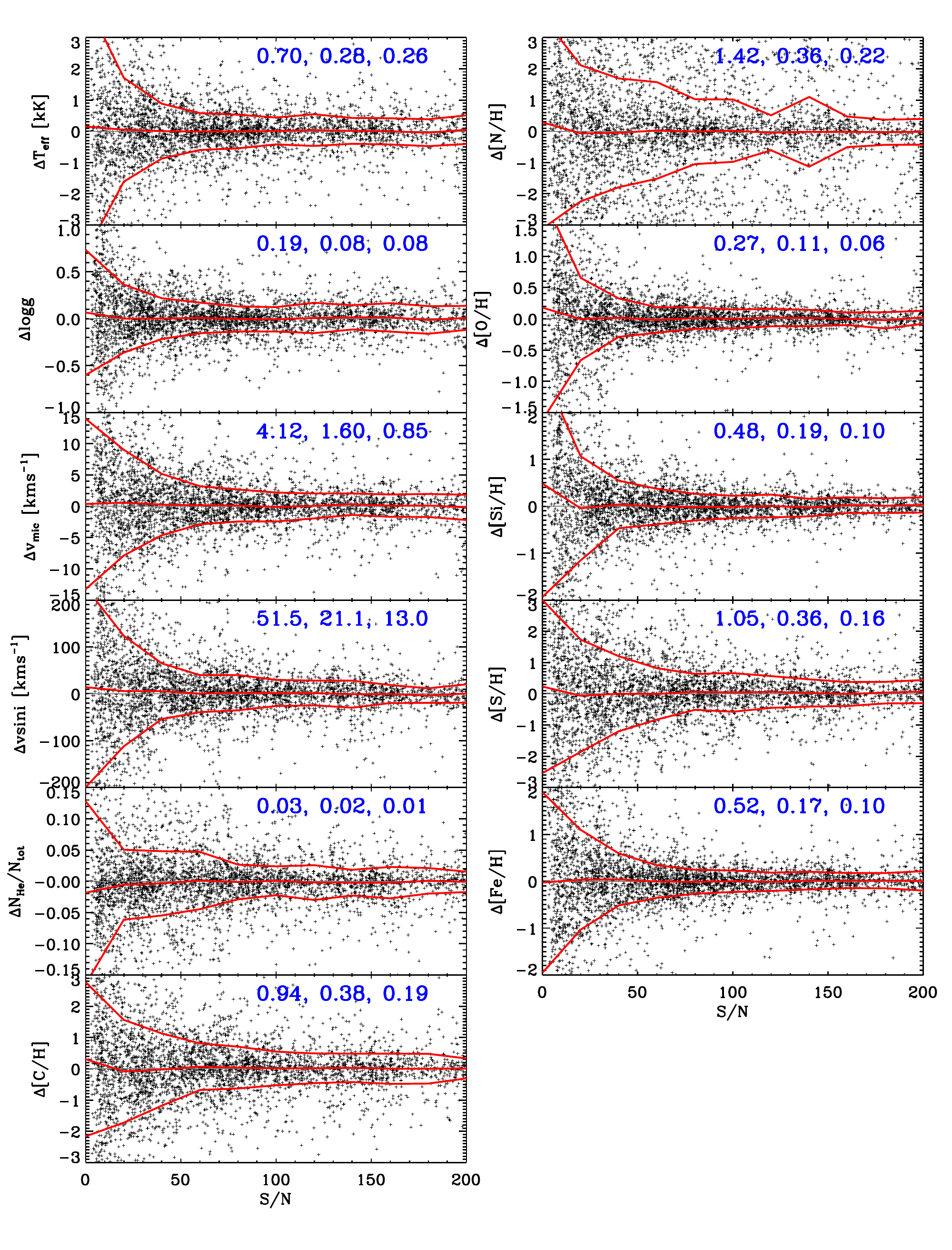}
\caption{Label differences between repeat observations as a function of S/N for stars with $\teff>10,000$\,K. Only repeat observations carried out in different observation nights and with S/N differences less than 20 are adopted. The x-axis shows the mean S/N of the repeat observations. The solid lines in red show the median differences, as well as the standard deviations at different S/N. The numbers marked in the figure are precision (the scatter divided by $\sqrt{2}$, i.e., assuming both stars contribute equally) for stars with $S/N<50$, $50<S/N<100$, and $S/N>100$, respectively.}
\label{fig:fig10}
\end{figure*}
About 30\% of the LAMOST spectra are repeat observations of common targets \citep[e.g.][]{Yuan2015, Xiang2017b}. The label estimates from these repeat observations can differ due to a number of reasons, such as the photon noise, the fiber-to-fiber and temporal variation of LSF, and imperfect calibration processes, e.g. flat-fielding, sky/scatter light subtraction etc. Comparing the label estimates from the repeat observations thus gives an estimate of the precision of the label estimates \footnote{Stellar variability could be an issue as well, but we do not expect it dominates over the measurement uncertainties for the majority of stars.}. To estimate the quality of our fit, we adopt only the repeat observations that have comparable ($<20\%$ difference) spectral S/N. The precision of the label is estimated through the square root of the dispersion between the repeat observations. 

The precision of our estimates as a function of S/N is shown in Fig.\,\ref{fig:fig10}. The Figure shows that, at S/N below 100, the photon noise dominates the spectral uncertainties, with the precision improve as 1/(S/N), as expected from the Cram\'er-Rao bound calculation. But the precision only gradually improve beyond $S/N>100$ because the instrument and calibration errors dominate the spectral uncertainties. In the case of $S/N<50$, the label estimates have a precision of 700\,K in \teff, 0.19\,dex in \logg, 4.1\,km/s in \vmic, 51.5\,km/s in $v{\rm sin}i$, 0.03 in $N_{\rm He}/N_{\rm tot}$, 0.27\,dex in [O/H], $\sim0.5$\,dex in [Si/H], and [Fe/H], $\gtrsim$1.0\,dex in [C/H], [N/H], and [S/H]. Whereas in the case of $S/N>100$, the uncertainties reduce to 260\,K in \teff, 0.08\,dex in \logg, 0.85\,km/s in \vmic, 13.0\,km/s in \vsini, 0.01 in \he, $\sim0.20$\,dex in [C/H], [N/H], and [S/H], $\lesssim$0.10\,dex in [O/H], [Si/H], and [Fe/H]. In general, these values for the case of $S/N \simeq 100$ are consistent with the CR bounds, within about a factor of 2. 

\subsection{Comparison with literature}
The LAMOST surveys mostly target stars fainter than 10\,mag in SDSS $r$-band, however most of the literature stars are brighter. As a consequence, there is a lack of a meaningful sample of reference hot stars that have well-established parameters and abundances from high-resolution, NLTE-based spectroscopy. For the current purpose, we compile our reference stars from two sources, the first is a collection of high-resolution, NLTE-based stellar labels of A-type and B-type stars from \citet{Przybilla2006}, \citet{Przybilla2010}, and \citet{Nieva2012}. These stars were not observed by the LAMOST survey but have available  high-resolution full-optical spectra from the ESO archive. We degrade them to the LAMOST resolution with the mean LSF to perform the label determination with the \hotpayne. Table\,\ref{table:table2} presents a summary of the sources of high-resolution spectra for our reference stars, with 11 stars in total. Their stellar parameters and abundances from high-resolultion, NLTE analysis in literature are presented in Table\,\ref{table:table3} and Table\,\ref{table:table4}, respectively. 

The other stars in our reference sample are slowly pulsating B-type (SPB) stars from \citet{Gebruers2021}. The \citet{Gebruers2021} sample consists 20 SPB stars in the $Kepler$ field, and with light curves from the $Kepler$ mission \citep{Borucki2010}. The stellar labels of these SPB stars are estimated from high-resolution spectra, utilizing LTE-based stellar model spectra \citep{Gebruers2021}. Among the 20 SPB stars of \citet{Gebruers2021}, we found 18 of them having LAMOST spectra from the LAMOST-$Kepler$ project \citep{DeCat2015, Fu2020}. Due to their intrinsic brightness, these LAMOST spectra all have $S/N>200$.  
\begin{table}
\centering
\caption{Sources of high-resolution optical spectra from ESO archives}
\label{table:table2}
\begin{tabular}{ccc}
\hline
 Star ID &  Wavelength range  & Instrument  \\
\hline
 HD149438 & 3800--6800\AA & HARPs  \\
 HD36512 & 3700--9000\AA & UVES \\
 HD36822 & 3700--9000\AA & UVES \\
 HD36960  & 3700--9000\AA & UVES \\
 HD172167 & 6800--9000\AA & UVES \\
     & 3700--4000\AA & UVES \\
    & 4000--6800\AA & \citet{Takeda2007} \\
HD34816  & 3800--9000\AA  & XSHOOTER  \\
HD35299 & 3700--9000\AA & XSHOOTER \\
HD87737 & 3700--9000\AA & XSHOOTER \\  
HD92207     & 3700--9000\AA & XSHOOTER \\
HD111613    & 3700--9000\AA & XSHOOTER \\
HD209008  & 3700--9000\AA & XSHOOTER \\
 \hline
\end{tabular}
\end{table} 

\begin{table*}
\centering
\caption{Stellar parameters of our reference stars from high-resolution, NLTE analysis$^1$}
\label{table:table3}
\begin{tabular}{cccccc}
\hline
 Star ID &  \teff  & \logg & \feh & \vmic & \vsini  \\
\hline
 HD111613 & $9150\pm150$ & $1.45\pm0.10$ & $-0.11\pm0.10$ & $7.0\pm1.0$ & $19.0\pm3.0$\\
 HD92207 & $9500\pm200$ & $1.20\pm0.10$ & $-0.16\pm0.07$ & $8.0\pm1.0$ & $30.0\pm5.0$\\
 HD172167 & $9550\pm150$ & $3.95\pm0.10$ & $-0.54\pm0.10$ & $2.0\pm0.5$ & $22.0\pm2.0$\\
 HD87737 & $9600\pm150$ & $2.00\pm0.15$ & $-0.07\pm0.10$ & $4.0\pm1.0$ & $0.0\pm3.0$\\
 HD209008 & $15800\pm200$ & $3.75\pm0.05$ & $0.03\pm0.08$ & $4.0\pm1.0$ & $15.0\pm3.0$\\
 HD35299 & $23500\pm300$ & $4.20\pm0.05$ & $0.03\pm0.10$ & $0.0\pm1.0$ & $8.0\pm1.0$\\
 HD36960 & $29000\pm300$ & $4.10\pm0.07$ & $-0.02\pm0.09$ & $4.0\pm1.0$ & $28.0\pm3.0$\\
 HD36822 & $30000\pm300$ & $4.05\pm0.10$ & $0.02\pm0.04$ & $8.0\pm1.0$ & $28.0\pm2.0$\\
 HD34816 & $30400\pm300$ & $4.30\pm0.05$ & $0.04\pm0.07$ & $4.0\pm1.0$ & $30.0\pm2.0$\\
 HD149438 & $32000\pm300$ & $4.30\pm0.05$ & $0.04\pm0.09$ & $5.0\pm1.0$ & $4.0\pm1.0$  \\
 HD36512 & $33400\pm200$ & $4.30\pm0.05$ & $0.03\pm0.03$ & $4.0\pm1.0$ & $20.0\pm2.0$\\
 \hline
\end{tabular}
\begin{tablenotes}
      \small
      \item{$^1$ The stellar parameters are from a collection of literature values in \citet{Przybilla2006}, \citet{Przybilla2010} and \citet{Nieva2012}}.
\end{tablenotes}
\end{table*} 

\begin{table*}
\centering
\caption{Elemental abundances of our reference stars from high-resolution, NLTE analysis$^1$}
\label{table:table4}
\begin{tabular}{ccccccc}
\hline
 Star ID &  \he  & [C/H] & [N/H] & [O/H] & [Si/H] & [S/H]  \\
\hline
 HD111613 & $0.105\pm0.020$ & $-0.26\pm0.07$ & $0.57\pm0.10$ & $0.01\pm0.04$ & $-0.06\pm0.29$ & $-0.05\pm0.08$ \\
 HD92207 & $0.12\pm0.02$ & $-0.10\pm0.20$ & $0.42\pm0.04$ & $0.10\pm0.07$ & $-0.18\pm0.05$ & $0.00\pm0.08$ \\
 HD172167 & $0.090\pm0.010$ & $-0.20\pm0.11$ & $-0.14\pm0.06$ & $-0.12\pm0.05$ & $-0.57\pm0.05$ & $-0.06\pm0.03$ \\
 HD87737 & $0.13\pm0.02$ & $-0.41\pm0.10$ & $0.58\pm0.09$ & $0.09\pm0.05$ & $0.07\pm0.19$ & $0.03\pm0.07$ \\
 HD209008 & 0.086$^2$ & $-0.10\pm0.09$ & $-0.03\pm0.11$ & $0.11\pm0.11$ & $-0.09\pm0.04$ & - \\
 HD35299 & 0.086$^2$ & $-0.08\pm0.09$ & $-0.01\pm0.08$ & $0.15\pm0.09$ & $0.05\pm0.05$ & - \\
 HD36960 & 0.086$^2$ & $-0.08\pm0.09$ & $-0.11\pm0.11$ & $-0.02\pm0.08$ & $0.05\pm0.07$ & - \\
 HD36822 & 0.086$^2$ & $-0.15\pm0.14$ & $0.09\pm0.10$ & $-0.01\pm0.10$ & $0.05\pm0.07$ & - \\
 HD34816 & 0.086$^2$ & $-0.05\pm0.05$ & $-0.02\pm0.15$ & $0.02\pm0.09$ & $0.03\pm0.06$ & - \\
 HD149438 & $0.089\pm0.009$ & $-0.13\pm0.12$ & $0.33\pm0.12$ & $0.08\pm0.08$ & $0.01\pm0.06$ & - \\
 HD36512 & 0.086$^2$ & $-0.08\pm0.14$ & $-0.04\pm0.11$ & $0.06\pm0.09$ & $0.03\pm0.07$ & - \\
 \hline
\end{tabular}
\begin{tablenotes}
      \small
      \item{$^1$ The stellar abundances are from a collection of literature values in \citet{Przybilla2006}, \citet{Przybilla2010} and \citet{Nieva2012}.}
      \item{$^2$ A fixed He abundance of 0.086 is adopted for the elemental abundances determination.}
\end{tablenotes}
\end{table*} 

\begin{figure*}[hbt!]
\centering
\includegraphics[width=1.0\textwidth]{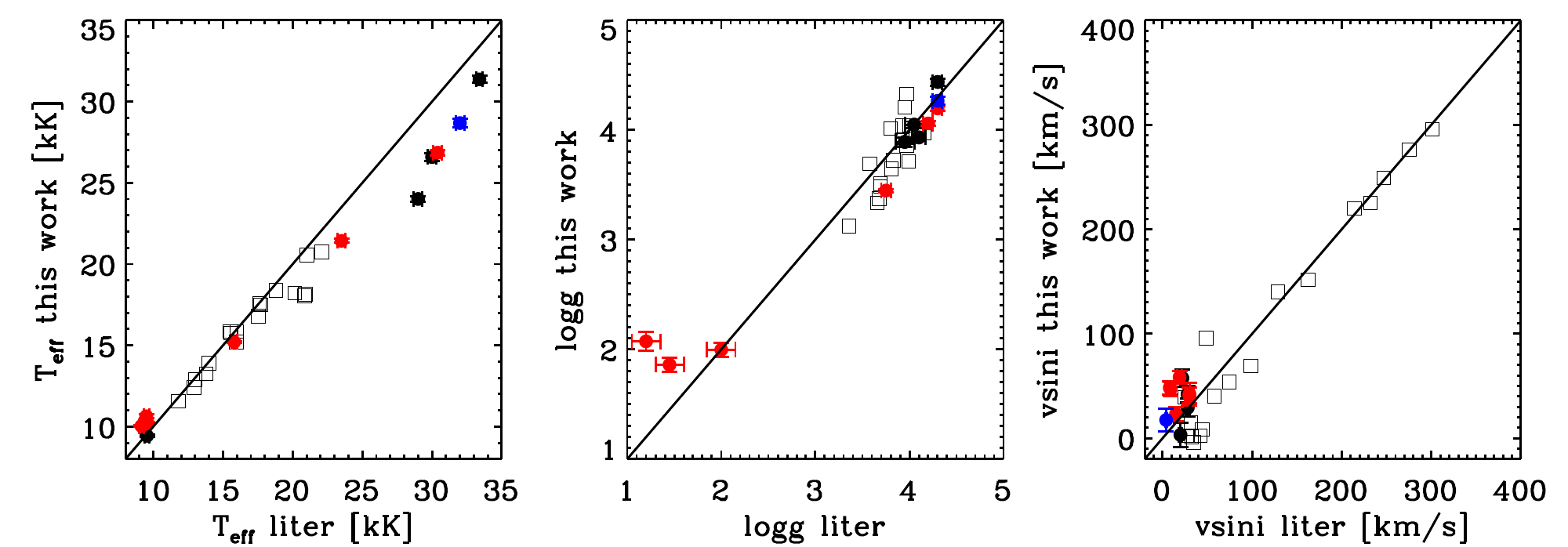}
\caption{Comparison of labels between the \hotpayne estimates and high-resolution spectroscopic results for our reference stars. For all the dots in black, blue, and red, the spectra are collected from the ESO archive (Table\,\ref{table:table2}), and degraded to the LAMOST resolution for stellar label determination with the \hotpayne. The dots in black, red, and blue represent reference stars whose spectra are from UVES, XSHOOTER, and HARPS spectrographs, respectively. For these stars, the literature labels, which are based on high-resolution, NLTE modeling, are adopted from \citet{Przybilla2006}, \citet{Przybilla2010} and \citet{Nieva2012}. Open squares represent the SPB stars of \citet{Gebruers2021}, which have LAMOST observational counterparts. The high-resolution (literature) spectroscopic labels of these SPB stars are derived using LTE-based model spectra. The \hotpayne  estimates show good agreements with the literature values For stars with $\teff>25,000$\,K, the LTE-based estimation with the \hotpayne underestimate the \teff due to NLTE effect. Similarly, the \logg for supergiants ($\logg<2$) are overestimated due to the differences between the LTE and NLTE modeling.}
\label{fig:fig11}%
\end{figure*}

\begin{figure*}[hbt!]
\centering
\includegraphics[width=1.0\textwidth]{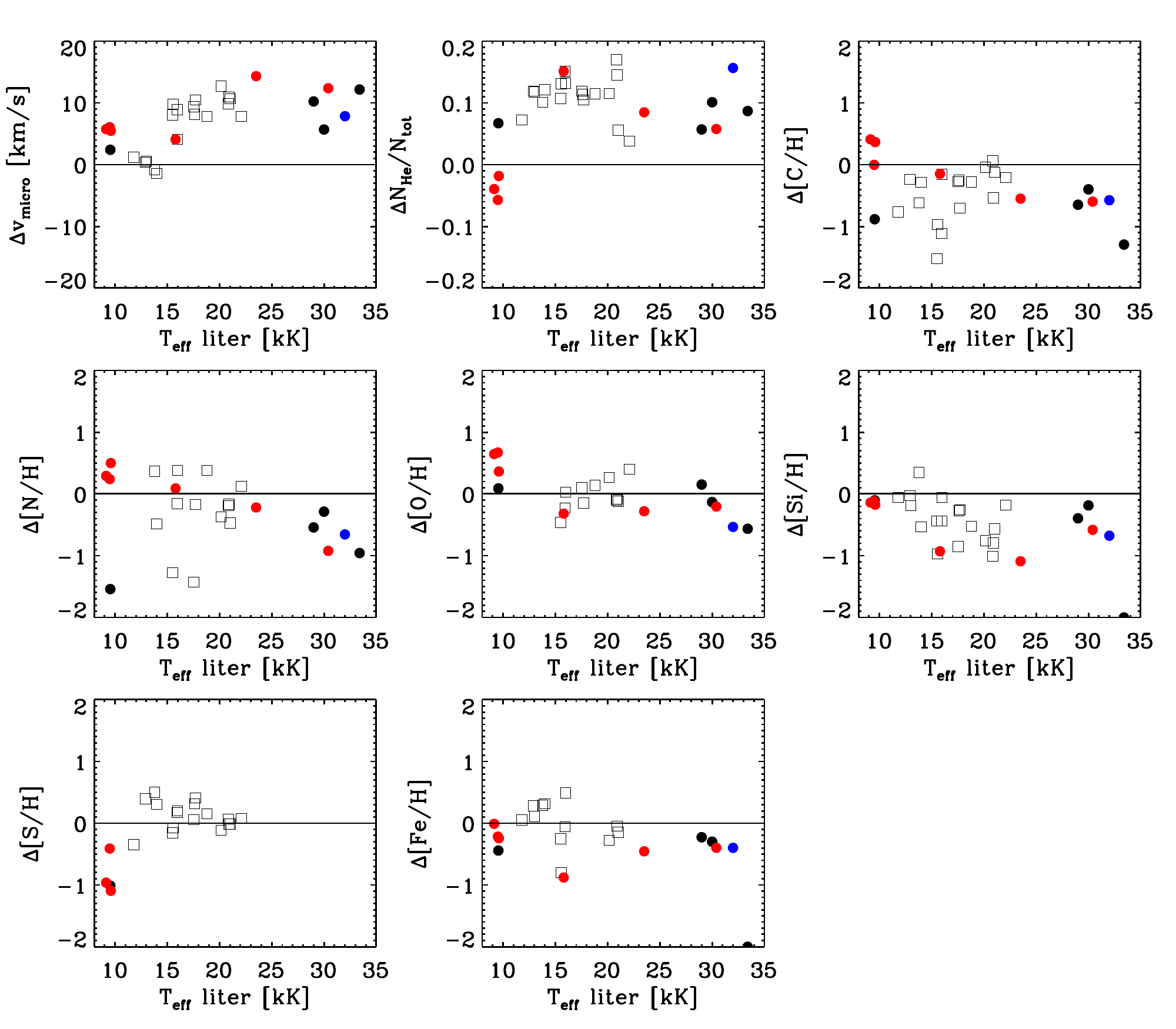}
\caption{Difference (LAMOST - literature) between stellar labels estimated from the LAMOST $R\sim1800$ spectra with\hotpayne and the literature values deduced from high-resolution spectroscopy for the reference stars, as a function of \teff. The symbols are defined the same way as those in Fig.\,\ref{fig:fig12}. Some of the differences are likely dominated by the differences between the LTE (this study) and NLTE modeling. See text for details.}
\label{fig:fig12}%
\end{figure*}

Fig.\,\ref{fig:fig11} presents an one-to-one comparison of the labels estimates from \hotpayne with their reference values for \teff, \logg, and \vsini. The comparison suggests that our estimates for these labels are in decent agreement with the literature values, except for the \teff of stars hotter than 25,000\,K. For these hot stars, our \teff estimates are systematically lower than the literature values by about 3000\,K, due to the NLTE effects. Since the NLTE models generally predict stronger lines than LTE models, as a result, the LTE analysis favors a lower \teff for these hot stars (see Appendix). The \logg shows good agreement with our results, demonstrating the ability of \hotpayne to distinguish giants from dwarfs. A particularly interesting failure mode is the A-giant star, HD92207, which exhibits large deviation in \logg. For this star, \citet{Przybilla2006} gives an estimate of 1.2\,dex for \logg while our estimate is 2.0\,dex. This discrepancy is caused by NLTE effects, which are particularly strong for supergiants \citep[e.g.][]{Kudritzki1976,Kudritzki1979}. The \vsini shows good agreement with literature, particularly for stars rotating faster than 100\,km/s, demonstrating the ability of \hotpayne to distinguish fast rotating stars from slow rotating stars. 

Fig.\,\ref{fig:fig12} illustrates a similar comparison but \vmic, \he, and [$X$/H]. The differences between the \hotpayne estimates and the literature values are plotted as a function of \teff  of the stars. For stars hotter than 15,000\,K, our estimates of \vmic are systematically larger than the literature values by up to 10\,km/s, due to NLTE effect, which is more prominent for the hotter stars (see Appendix). At cooler end ($\teff<15,000$\,K), our \vmic estimates for dwarfs are in decent agreement with literature values. For example, for Vega (HD172167), the literature \vmic is 2\,km/s, and we obtain 4.4\,km/s. Whereas for the A-type supergiants, our estimates are higher than literature values, again because of NLTE effect, which is more prominent for supergiants than for dwarfs. Although not shown here, for the LAMOST stars with $\teff<9000$\,K, our results exhibit a typical \vmic of $\sim2$\,km/s, which is in line with expectation \citep[e.g.][]{Gebran2014}. The \he is systematically overestimated by $\sim0.10$, except for the A-type supergiants. This is  due to NLTE effect, as explained above, NLTE models have deeper He lines than LTE spectra, which in turn favor lower\he estimates. 

As for the abundances, the comparison reveals a \teff-dependent systematic trend in our abundance estimates, largely a consequence of NLTE effects. In particular, for stars with $\teff>25,000$\,K, our estimates are generally lower than the reference values by 0.3 for [Fe/H], 0.5--1.0\,dex for [C/H], [Si/H], and [S/H]. Note that for HD36512 (33,400K, in black, and barely visible in the plot), our estimates of [Si/H] and [Fe/H] are are erroneously low ($<-2$\,dex). The exact cause is hard to be determined, but it is likely due to the imperfect fringing subtraction of the UVES spectra, which is the most prominent for HD36152. 
For stars of $11,000<\teff<25,000$\,K, our [Fe/H] estimates are decent agreements with \citet{Gebruers2021}, but systematically higher than \citet{Przybilla2010, Nieva2012}. Both ours and \citet{Gebruers2021} are based on LTE models, and the good agreement validates our method, further elucidating that the differences with the other studies are largely due to the NLTE effect. We have only one star with NLTE abundances at $\teff\sim15,000$\,K. Beside NLTE effects, imperfections of line lists may have also played a role for the difference. Particularly, for C, we have adopted the older line list of \citet{Jorgensen1996}, which may need to be updated with more recent line lists \citep[e.g.][]{Masseron2014}. The [O/H] seems to be in decent agreement with literature, except for the A-type supergiants and the stars with $\teff>30,000$\,K, both of which may have suffered strong NLTE effects. The decent agreement of [O/H] with literature is not surprising because the Oxygen has the most prominent features in the spectra (see Appendix). Nonetheless, it seems that even for [O/H], there can a systematic uncertainty at a level of 0.2\,dex from NLTE effects.       

\section{Results and discussion}\label{sec:results_and_discussion}
\subsection{The final sample}
We apply the \hotpayne to our 1,160,000 LAMOST hot star candidate spectra. A considerable fraction of these candidates are found to be cool stars with $\teff<7000$\,K, a consequence of contamination from our target selection. Since our Kurucz spectral model grid covers only $\teff>7500$\,K (Table\,\ref{table:table1}), the labels for stars with $\teff<7500$\,K are estimated with the \hotpayne via extrapolation. We therefore discard all stars with $\teff<7000$\,K in our sample. Note that we choose to keep the stars with $7000<\teff\lesssim7500$\,K, because this regime was poorly investigated by previous efforts of label estimates for FGK stellar sample \citep[e.g.][]{Xiang2019}, and our labels might be useful for the general purpose.

We discard results from spectra with $S/N<5$ from our final catalog to ensure precision of the label estimates. Furthermore, we found that the \teff estimates for a considerable number of ($\sim5\%$) intrinsically cool ($\teff<7000$\,K) stars are erroneously estimated to be hotter than 7000\,K due to their low spectral S/N. We also discard these stars from our sample. These outliers are identified based on their dereddened colors (see Section 5.3 for the derivation of dereddened colors) and effective temperature derived with the data-driven Payne \citep[DD-Payne;][]{Xiang2019}, which provides independent temperature estimates for FGK stars of $\teff<7000$\,K with a data-driven approach, taking stellar labels from the high-resolution surveys (APOGEE and Galah) as training sets. Specifically, if a star has a dereddened Gaia BP-RP color redder than 0.4 ($(BP-RP)_0>0.4$), and simultaneously, the DD-Payne gives $\teff<7000$\,K, we deem the star to be an intrinsically cool one but the current work erroneously gives a high temperature. This criterion is applied to all our sample stars except for those from the OB star sample of \citet{LiuZ2019}, as the latter contains mostly intrinsic hot stars. Ultimately, these criteria leave 332,172 unique stars with 454,693 measurements (spectra) in our final hot star sample.

While we have shown that our abundances are internally consistent through repeat spectra (Fig.~\ref{fig:fig10}), the comparison with the literature values reveals that NLTE effects can remain critical, especially for the hot stars, we opt not present all the abundance estimates in our final parameter catalog, to avoid any misuse of our catalog.  Instead, we only present the \feh and [Si/H] as the metallicity and alpha-enhancement indicators. As we will show in Section\,5.5, despite the zero-point offset due to the NLTE systematics, these estimates (e.g., [Si/H]) are still informative for chemically peculiar stars. Other abundances are available upon request. We also note that we are in the process of applying the same method to NLTE models, and will release the full NLTE abundances in the coming study.

\subsection{Flagging bad fits}
As we are working with a vast set of spectra, collected by a complex instrument system (16 spectrographs, 32 CCDs, 4000 fibers) of LAMOST \citep{Cui2012}, inevitably there will be some ``bad" spectra with errorneous label estimates, either due to intrinsic peculiar properties, e.g. strong emission-line objects, or due to data reduction artefacts \citep{Xiang2021}. To ensure the robustness of our results, we first identify such cases by flagging results with unusually large $\chi^2$ in the spectral fitting. We quantify the median and dispersion\footnote{The dispersion is defined as a resistant estimate of the standard deviation, adopting the $robust\_sigma.pro$ in the IDL astronomical library. See e.g. \citet{Hoaglin1983} for the method of resistant estimate of the dispersion of a distribution.} of the reduced $\chi^2$ as a function of \teff and S/N. We then define a flag as `chi2ratio' in the catalog to describe the deviation of the reduced $\chi^2$ for individual stars from the median value of stars with similar S/N and \teff, divided by the dispersion. We model the reduced $\chi^2$ for both the median and dispersion as a function of \teff and S/N with a 3-order polynomial. In the analysis below, we discard the stars with reduced $\chi^2$ larger than $10\sigma$ from the median value (`chi2ratio$>$10'). We do not adopt a constant $\chi^2$ cut because the typical $\chi^2$ of the spectral fitting is found to vary with the \teff and S/N of the spectra, due to both imperfect flux uncertainty estimates of the LAMOST spectra, i.e., overestimating the flux uncertainty for low S/N stars or underestimating the flux uncertainty for high S/N stars. Either case, these can lead to a S/N dependent $\chi^2$ distribution. Furthermore,  since the model accuracy varies as a function of \teff, for instance, the Kurucz spectra are expected to be less accurate for very hot ($\teff>25,000$\,K) stars, modeling `chi2ratio' as a function of \teff is to ensure that the $\chi^2$ criteria is adjusted according to the model accuracy.

\subsection{\teff and \logg}
\begin{figure}[hbt!]
\centering
\includegraphics[width=0.48\textwidth]{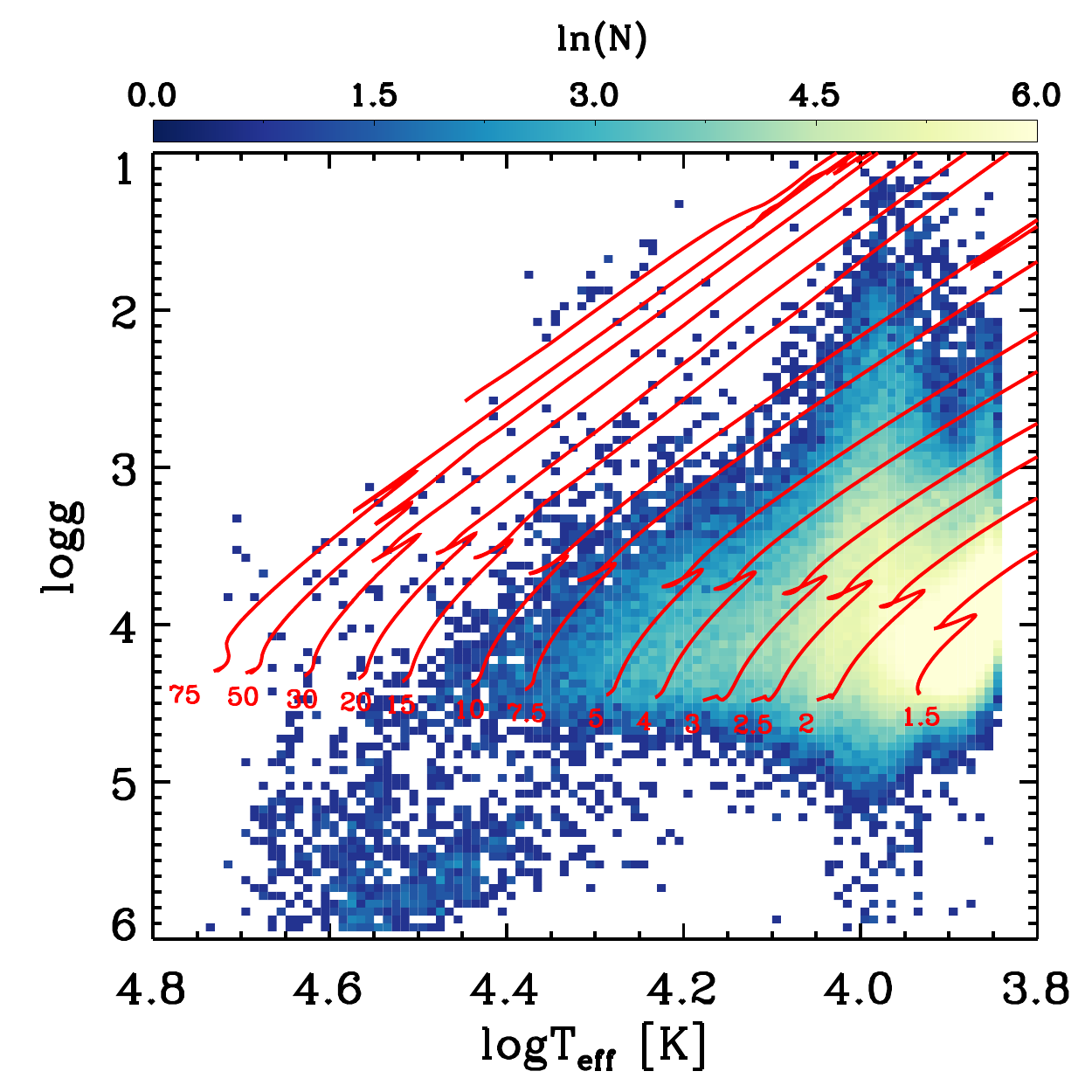}
\caption{Stellar number density distribution in the Kiel (\teff--\logg) diagram. Only 211,245 stars with $S/N>30$ are shown. Also shown in the figures are the MIST stellar evolution tracks \citep{Paxton2011, Choi2016} with $\feh=-0.5$ and $V/V_{\rm c} = 0.4$. The initial stellar masses of the tracks are marked in the Figure (in unit of $M_\odot$). The stars with $\logg>5$ and $\log\teff>4.2$ are hot subdwarfs. The vertical stripe at $\log\teff\sim4.0$ is an artefact for chemically peculiar stars (e.g. Ap/Bp stars, Am stars) whose \logg have been underestimated as a consequence of masking the Hydrogen lines for label determination.}
\label{fig:fig13}%
\end{figure}
Fig.\,\ref{fig:fig13} presents the distribution of our sample stars with $S/N>20$ in the \teff--\logg diagram over plotted with the MIST stellar evolutionary tracks \citep{Paxton2011, Choi2016}, the numbers marked in red are the initial mass of the the stellar evolutionary tracks. The majority of the sample stars have relatively cool temperatures, corresponding to stellar masses of 1.5--7.5\msun. While there are also a number of stars whose locations in the diagram are consistent with stellar mass higher than 20\msun. At the high temperature end, there are a group of stars with $\logg\gtrsim5$, which are outside the coverage of the stellar evolution tracks. These stars are mostly hot subdwarfs (e.g. sdBs, sdOs), whose origin is still a research topic \citep[e.g.][]{Webbink1984, Han2002, Han2003, Lanz2004, MillerBertolami2008, Heber2009, Justham2011, Zhang2012}. Hot subdwarfs  have been extensively identified from the LAMOST database by \citet{Luo2016, Luo2019, Luo2020}. The positions of the hot subdwarfs in the \teff--\logg plane are consistent with \citet[][e.g. see their Fig.\,4]{Luo2016}, which is remarkable since our \logg estimates for these stars are obtained via extrapolation of \hotpayne. It is also possible that a minor number of these stars might be contamination of white dwarfs. We caution about the usage of our stellar labels for hot subdwarfs and white dwarfs, as they are estimated with model spectra extrapolated with the \hotpayne.

At $\teff\sim10,000$\,K ($\log\teff\sim4$), there is a vertical stripe of stars that extend to low \logg values ($<2$). As we will further discuss in Section 5.5, this is an artefact due to the existence of a large number of chemically peculiar stars, such as ApBp stars and Am stars. The \logg of these stars may have been significantly underestimated as a consequence of masking the Hydrogen lines for label determination for these early A-type and late B-type stars (Section\,4.4). We note that such stars are prevalence in the hot stars regime; it is suggested that a high fraction of A stars could be chemically peculiar \citep[e.g.][]{Gray2016, Qin2019, Xiang2020}, and this is why their high number density constitutes a prominent stripe in the \teff$-$\logg plane. Such artefact reveals one of the limitations of the current model, which tends to yield incorrect \logg for these chemically peculiar stars. However, as will show in Section 5.5, the abundance estimates of these stars are nonetheless robust enough, and thus these stars can be easily flagged through their abundances estimates. 

\begin{figure*}[hbt!]
\centering
\includegraphics[width=1.0\textwidth]{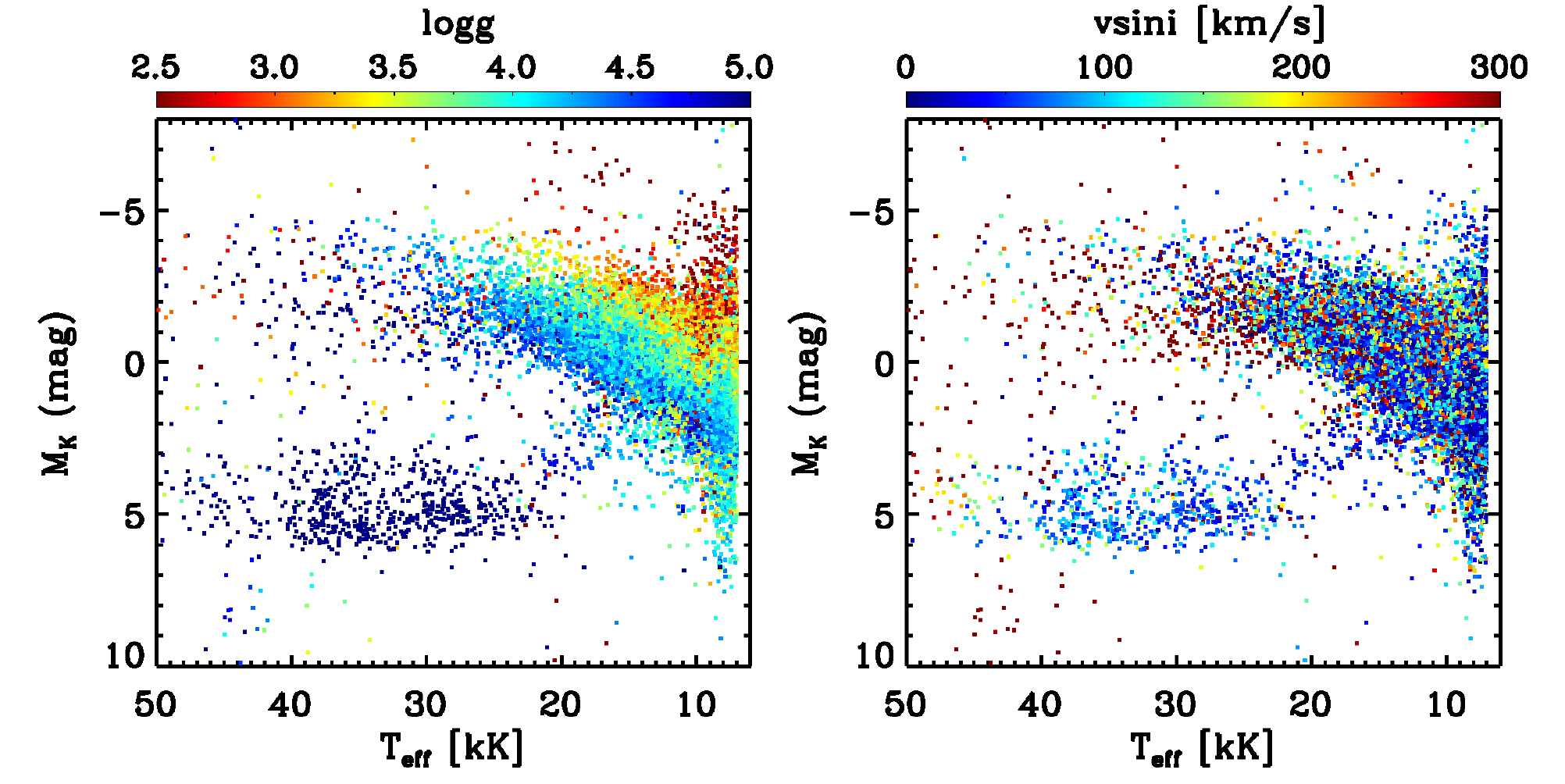}
\caption{Stellar distribution in the \teff--\mk diagram color-coded by \logg (left) and \vsini (right). Only 194,724 stars with good spectral S/N ($S/N>30$) and good parallax ($\omega/\sigma_\omega>5$) are shown. The expected correlation between \logg with \mk at a given \teff demonstrates the robustness of the spectroscopic \logg estimates. The stars with $\teff\gtrsim20,000$\,K and $\mk\sim5$\,mag are hot subdwarfs. In the left panel, a maximal \logg value of 5 is displayed in the color bar, all stars with $\logg>5$ are shown with the same color as $\logg=5$ (dark blue). }
\label{fig:fig14}%
\end{figure*}
About 98\% of our sample stars have parallax measurements from the Gaia eDR3 \citep{Brown2021}, which allow us to look into their luminosity (absolute magnitude). We derive their absolute magnitude in the 2MASS K-band \citep{Skrutskie2006} using the distance modulus, adopting the Gaia eDR3 distance of \citet{Bailer-Jones2021}. The interstellar extinction is corrected using the reddening $E(B-V)$ interpolated from the 3D reddening map of \citet{Green2019} and a K-band extinction coefficient of 0.34 \cite[e.g.][]{Yuan2013}. Fig.\,\ref{fig:fig14} presents the stellar distribution in the \teff--\mk diagram for a subset of 195,879 stars with good spectral S/N ($S/N>30$) and good parallax ($\omega/\sigma_\omega>5$). The figure illustrates a good correlation between \mk and \logg, suggesting our \logg estimates are robust for the overall sample stars. The figure also illustrates that at a given \teff, the \vsini exhibits little correlation with \mk. Finally, while relatively cool stars exhibit a wide range of \vsini values, the majority of hot stars with $\teff\gtrsim25,000$\,K tend to exhibit large \vsini values (except for hot subdwarfs). 

\begin{figure*}[hbt!]
\centering
\includegraphics[width=1.0\textwidth]{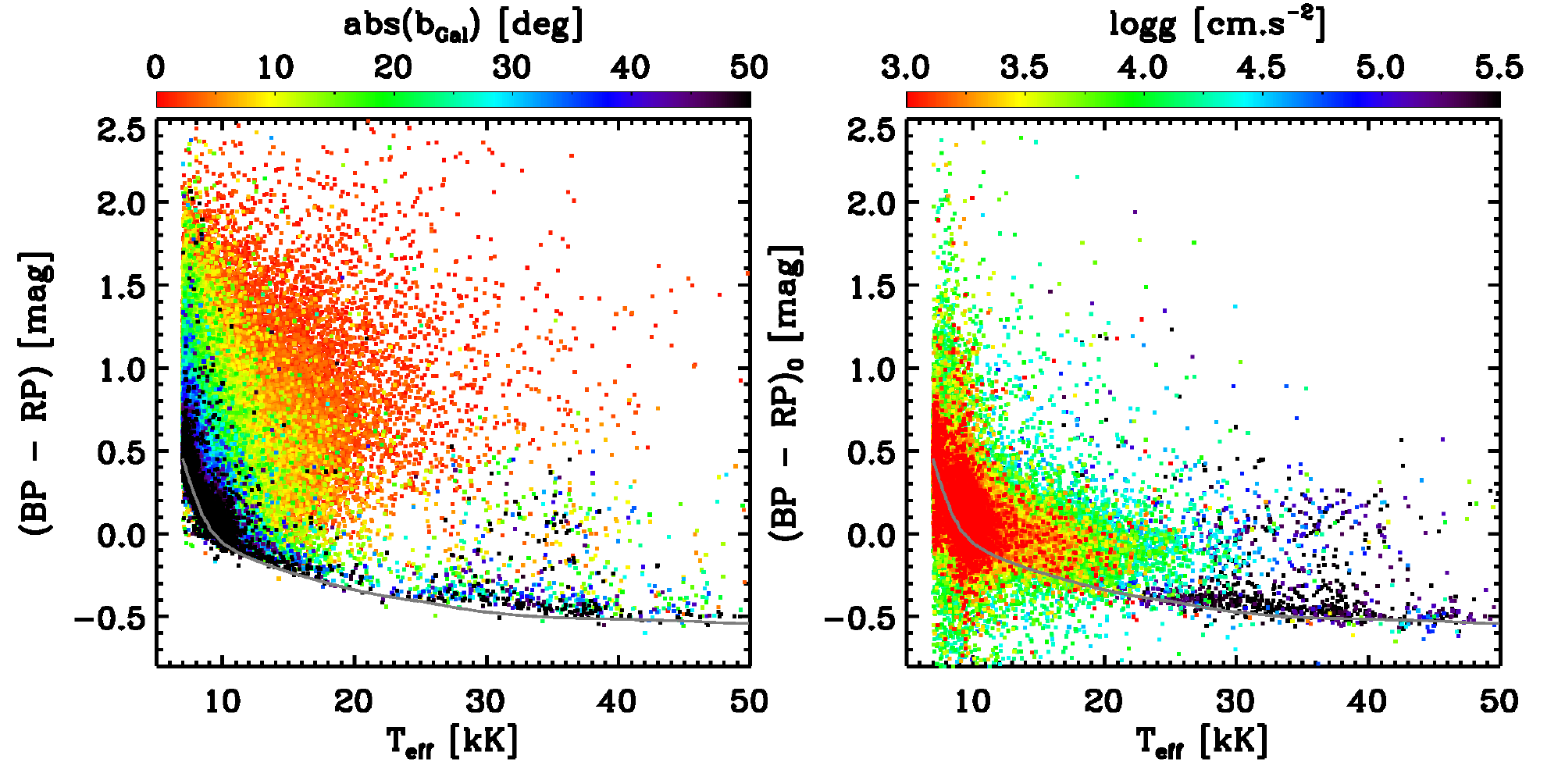}
\caption{Observed and de-reddened (BP - RP) color as a function of effective temperature. Only stars with $S/N_{spec}>30$ are shown.  The left panel shows the observed (BP - RP) color {\it vs.} $\teff$, with the stars color-coded by Galactic latitude as a rough proxy for the expected dust reddening. The grey line delineates the MIST isochrones's prediction for $(BP-RP)$ color as a function of \teff for zero-age main-sequence stars with initial $\feh=0$. This panel illustrates that our sample stars in low Galactic latitudes suffer from large reddening effect, leading to much redder colors than stars at high Galactic latitudes. The right panel shows (BP - RP)$_0$, the dereddened color, based on the 3D reddening map of \citet{Green2019} with the stars are sorted and color-coded by \logg. The panel illustrates that such de-reddened colors qualitatively agree with the isochrones, but the significant spread in color at any given temperature implies that the photometric colors alone would be too imprecise to estimate $\teff$ robustly for stars above 10,000~K. The very hot stars with little reddening and very high $\logg$ are presumably nearby subdwarf B stars of modest luminosity (see Fig.~\ref{fig:fig14}).  }
\label{fig:fig15}%
\end{figure*}
In Fig.\,\ref{fig:fig15}, we show the $(BP - RP)$ color of Gaia eDR3 as a function of our \teff estimates for stars with $S/N>30$. There is a clear trend of $(BP - RP)$ with Galactic latitude, which is due to the reddening effect. For stars at high Galactic latitude ($|b|>30^\circ$), there is a clear relation between \teff and $(BP - RP)$, which is consistent well with the \teff-dependent trajectory of synthetic color from the MIST isochrones. Stars at low Galactic latitude exhibit a much broader $(BP - RP)$ distribution, due to their large reddening effect. The right panel of Fig.\,\ref{fig:fig15} presents a similar results but correcting the color for interstellar reddening $(BP - RP)_0$  estimated using the 3D reddening map of \citet{Green2019} and a total-to-selective extinction coefficient $R_{BP}=3.24$ and $R_{RP}=1.91$ \citep{Chen2019}. After dereddening, the colors of the stars agree approximately with the expectation from the MIST isochrones. Interestingly, even after correcting for the reddening, a large fraction of hot stars with $\teff\gtrsim25,000$\,K appear to be too red, with $(BP-RP)_0\sim0.0$. These stars are likely either associated with star formation regions (for young stars) or having dust envelopes (for hot subdwarfs) as such stars suffer (local) larger reddening than the \citet{Green2019} 3D ``average'' dust map (see also \citet{Xiang2021}). The NLTE effect  should push the temperature even higher than the current estimates and would only further exacerbate this discrepancy.

\subsection{\vsini}
\begin{figure*}[hbt!]
\centering
\includegraphics[width=0.9\textwidth]{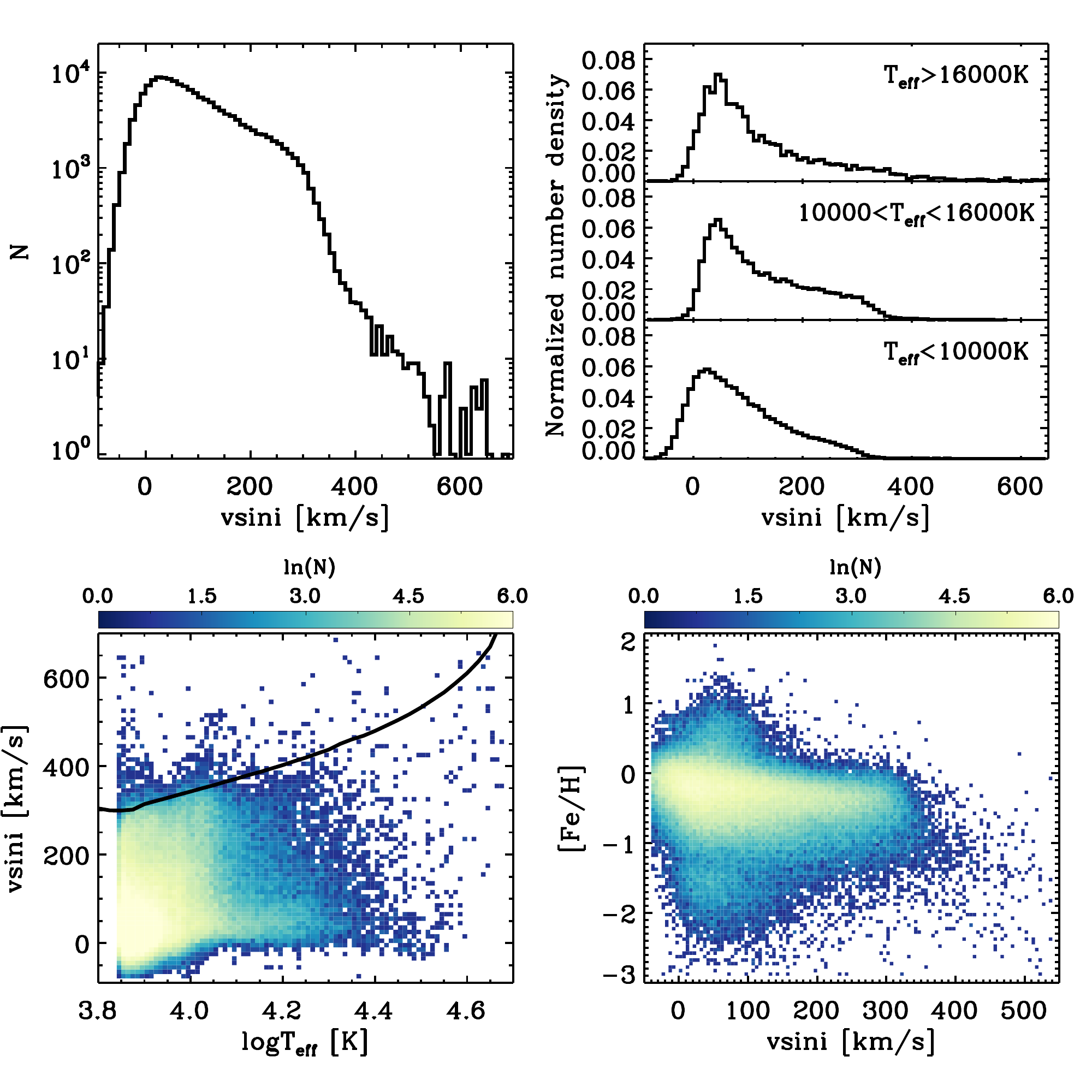}
\caption{Distribution of \vsini. Top left: histogram of \vsini for the overall sample in this work.  Only results from stars with $S/N>50$ and $\logg<5$ are shown.  Top right: histogram of \vsini for stars in different \teff range, as marked in the figure. Bottom left: the stellar number density distribution in the \teff--\vsini plane. The solid line delineates the critical rotation velocity $V_{\rm c}$ as a function of \teff, derived based on the zero-age main-sequence MIST isochrones \citep{Choi2016}. Bottom right: the stellar number density distribution in the \feh--\vsini plane. Since we do not impose a prior limit on \vsini, our fit allows for  negative \vsini values. A negative \vsini simply means that the star has a small rotation velocity, and the distribution at the negative tail serves as a proxy for the measurement uncertainties.} 
\label{fig:fig16}%
\end{figure*}
The \vsini of our sample stars with $S/N>50$ and $\logg<5$ are presented in Fig.\,\ref{fig:fig16}. Here we adopt stricter S/N cut than above for the \teff-\logg plot ($S/N>30$) to ensure the robustness of the \vsini for stars presented in the Figure. The \vsini estimates spread a broad range of values, from no rotation to faster than 600\,km/s. Note that our \vsini estimates could have negative values from the extrapolation of the model spectra with the \hotpayne. A negative \vsini means that the star has small rotation. In the final catalog, we have set all negative \vsini estimates to be 0\,km/s, but in Fig.\,\ref{fig:fig16} we keep the negative estimated values as they can serve as a good proxy for the uncertainties of the \vsini estimates. The extent of the negative tail suggests that, at least for the low \vsini, the measurement uncertainties are of the order of 30\,km/s. The \vsini distribution peaks at $\sim30$\,km/s, with a significant high-velocity tail extending to $\sim300$\,km/s, above which the number density free falls. This is qualitatively consistent with previous studies based on high-resolution spectroscopy, which have suggested that most of the early-type stars in nearby Galactic disk rotate slower than 300\,km/s, although a minor number of them can rotate faster than 400\,km/s \citep[e.g.][]{Abt2002, Huang2010, Braganca2012, Zorec2012, Simon-Diaz2014, LiGW2020}. 

Our results show that below 300\,km/s, the stellar number density exhibits a descending power-law. \citet{Abt2002} argued that the \vsini distribution of late B-type (B8-B9.5 III-V) stars exhibits double peaks, contributed by a set of slow rotating chemically peculiar stars and a set of fast rotating normal stars \citep[see also e.g.][]{Dufton2013}. We examine this idea via the \vsini distribution for our sample stars in different \teff bins. The middle of the top right panels of Fig.\,\ref{fig:fig16} show the results for stars with $10,000<\teff<16,000$\,K, which are composed of late B-type stars. Our results show that there is a prominent peak of slow rotating stars at $\vsini\sim30$\,km/s. Although there is indeed a significant tail of fast rotating stars (up to $\vsini\sim300$\,km/s), there is no clear double-peak feature. The reason for the difference is unclear yet, but we note that previous works usually have small sample size, typically a couple of hundred stars, while our sample contains 23,090 stars, about 50 times larger. Nonetheless, independent of the temperature, our results show that slow rotators ($\lesssim100$\,km/s) dominate over fast rotators for all stellar types from 7000\,K ($\log\teff\sim3.85$) to $25,000$\,K ($\log\teff\sim4.4$) (see the bottom left panel).  

The bottom left panel of Fig.\,\ref{fig:fig16} illustrates that at the high-velocity end, the stellar density exhibits a sharp cutoff, and the cutoff velocity increases with \teff. The cutoff is particularly prominent for stars with $\teff<20,000$\,K ($\log\teff\sim4.3$), above which the cutoff becomes less well defined due to the sparsity of stars. Theoretically, a star cannot be formed with infinity rotation velocity because the rotation will induce centrifugal force, which increases with the rotation velocity. Once the centrifugal force is larger than the gravity, the rotating disk that accretes materials and angular momentum to sustain the star formation will break up. This causes an upper limit of the rotation velocity, which is referred as the critical velocity, or the break-up velocity \citep[e.g.][]{Maeder2009}. The critical velocity can be expressed as, 
\begin{equation}
  V_c = \sqrt{\frac{2}{3}\frac{GM}{R_{\mathrm p}}},
\end{equation}
where $G$ is gravity constant, $M$ the stellar mass of a zero-age main sequence star, $R_{\mathrm p}$ the radius at the pole of the star. We assume that the $R_{\mathrm p}$ is approximately the same as the stellar radius in spherical models. In the bottom left panel of Fig.\,\ref{fig:fig16}, we show the critical velocity as a function of effective temperature. Here the critical velocity is derived using zero-age main-sequence (ZAMS) stars of the MIST stellar isochrones \citep{Paxton2011, Choi2016} with solar abundances. Technically, at each \teff, we define the main-sequence star with the maximal \logg as an approximation of the zero-age main-sequence star of that \teff.

It shows that, except for a few outliers, both the values and their \teff dependence of the observed \vsini border are consistent with the theoretical critical velocity, which increase from about 300\,km/s at $\teff=7000$\,K to $>600$\,km/s at $\teff=40,000$\,K ($\log\teff\sim4.6$). This good agreement suggests that there are indeed a considerable fraction of stars rotating with a rate close to the critical velocity. 

We note that a number of stars exhibit \vsini larger than the critical velocity. In principle, it is possible that a star might temporarily rotate faster than the critical velocity as a consequence of binary interaction \citep{de_Mink2013}. However, upon inspecting the spectra and intrinsic brightness of these outliers, we find that they are mainly composed of nearly equal-brightness binary stars that exhibit large velocity difference between the individual components thus their spectral lines are broader than single stars. In addition, these outliers also include stars whose \vsini are erroneously estimated due to a number of reasons, such as the occurrence of strong emission lines, which is particularly common for O-type stars, the erroneously spectral radial velocity correction, the contamination of subdwarfs and WDs, and the failure of the model spectra at the high-\teff end. For stars with $\teff\gtrsim25,000$\,K, the inaccurate line strength of the Kurucz LTE model spectra may cause problems for the \vsini estimates. For example, by matching the He line profile of LAMOST with NLTE model spectra, \citet{LiGW2020} suggest that the LAMOST J040643.69+542347.8 ($\teff\sim35,000$\,K) has a \vsini of $\sim$540\,km/s, while our estimate is $261\pm11$\,km/s, significantly underestimated due to the poor reproduction of the spectral line profile by our LTE model spectra at such high \teff. 

The bottom right panel of Fig.\,\ref{fig:fig16} presents the relation between \vsini and \feh. While there is no strong trend between \vsini and \feh for the majority of stars, a small group of stars with $\vsini\lesssim150$\,km/s exhibit high metallicity ($\feh\gtrsim0.2$). These stars are likely magnetic ApBp stars. We will further elaborate on this point in the next Section. At the metal-poor side, there are also an excess of stars with slow rotation velocities. These are found to be mostly halo blue horizontal branch (BHB) stars; due to their old ages, they are rotating much slower than the typical young hot stars in the disk.

\subsection{Abundance [Fe/H] and [Si/H]}
\begin{figure*}[hbt!]
\centering
\includegraphics[width=1.0\textwidth]{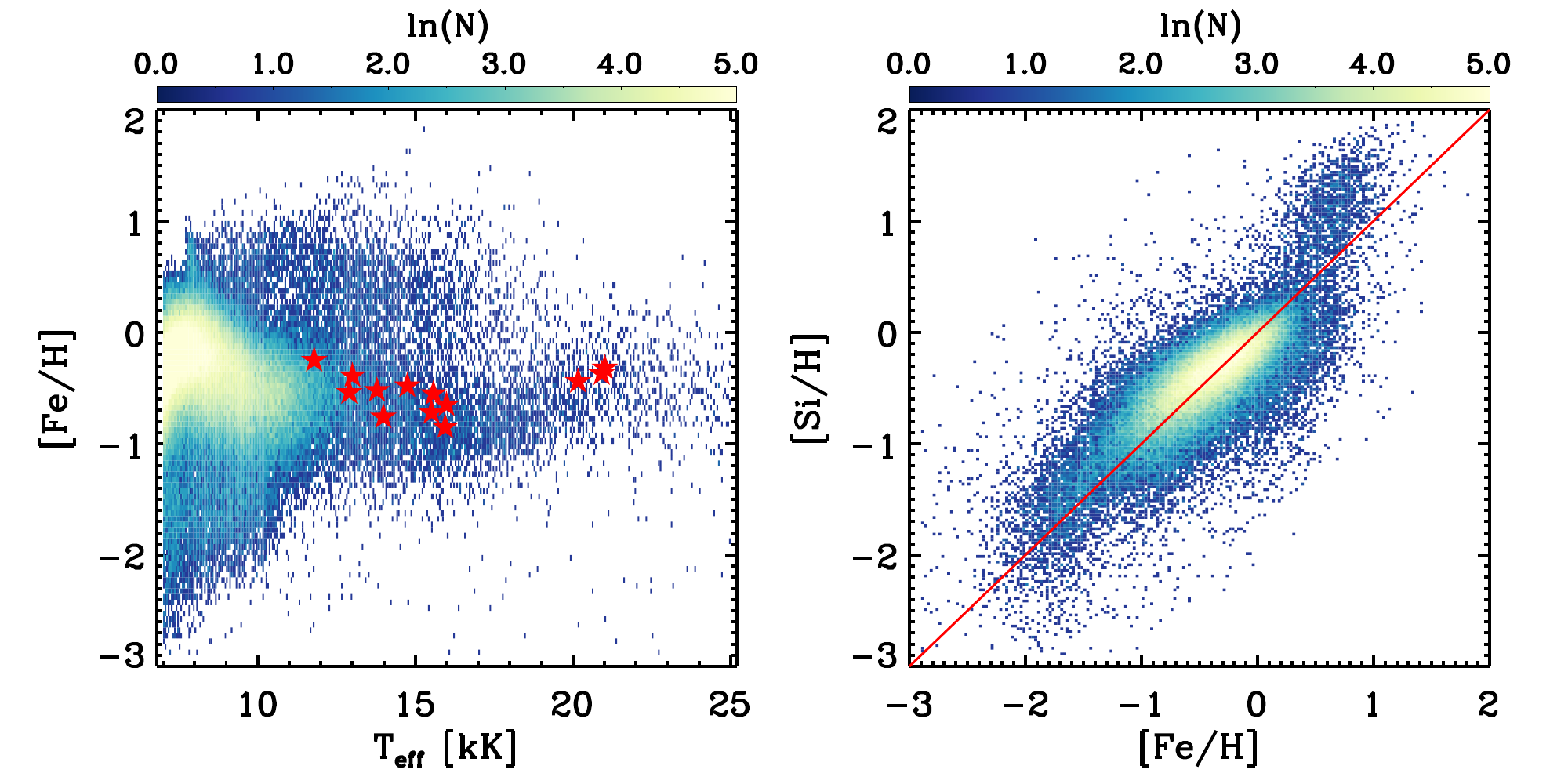}
\caption{Density distribution of our sample stars with $S/N>50$ in the \teff -- \feh (left) and \feh -- \sih (right) planes. In the left panel, the stars in red show the SPB stars of \citet{Gebruers2021}, with \feh from high-resolution spectroscopy (LTE-based). While the \teff-dependent trend of [Fe/H] distribution is likely due to NLTE effects, the figure clearly reveals the presence of chemically peculiar stars with super-solar \feh values, as well as halo BHB stars at the metal-poor end. The right panel shows good consistency between [Si/H] and [Fe/H], verifying the robustness of the determinations (regardless of the systematics due to NLTE). The group of stars with $\feh\sim0.7$\,dex and ${\rm [Si/H]}\gtrsim1$\,dex are  magnetic ApBp stars, indicated by their signature enhancement in Si\citep[e.g.][]{Michaud2015}. In the right panel, we omit stars with $\teff<8000$\,K because their \sih estimates are found to suffer from larger systematics.} 
\label{fig:fig17}%
\end{figure*}
In light of the large systematics and uncertainties of our LTE-based abundance estimates, we opt to present only the \feh and [Si/H] in our catalog, while leaving the other abundances to future work. The \feh is an essential metallicity indicator for relatively cool stars ($\teff\lesssim10000$\,K), both because the \feh of cool stars may spread a broad range of values as they may from different Galactic components (e.g. disk, halo), and because the \feh can be robustly estimated for cool stars owing to their prominent features in the spectra. Silicon is a representative $\alpha$-element. While the [Si/H] estimates may have large systematics (Fig.\,\ref{fig:fig12}), we expect their relative values are robust and informative for identifying chemical peculiars (Fig.\,\ref{fig:fig17}). 

The left panel of Fig.\,\ref{fig:fig17} shows the stellar distribution in the \teff--\feh plane. A prominent feature in the figure is the v-shape pattern of \feh -- typical \feh value for the bulk of stars decreases from $-0.2$\,dex at the cool end ($\teff\sim7000$\,K) to a minimal value of about $-0.7$\,dex at $\teff\sim17,000$\,K, beyond which the \feh slightly increases with \teff. This v-shape trend seems to be consistent with the high-resolution results of \citet{Gebruers2021}. The origin of this trend is unclear. It is possibly an artefact due to NLTE effects. Beyond this trend, there is a group of stars with super-solar metallicity at $\teff<20,000$\,K, which are clearly separated from the bulk of stars in the \feh-\teff plane. For many of these Fe-rich stars, the [Si/H] are also significantly enhanced (see the right panel of Fig.\,\ref{fig:fig17}). The high [Si/H] nature suggests these stars are likely chemically peculiar stars whose origin may be related to strong magnetic field, i.e., magnetic ApBp stars \citep[e.g.][]{Abt1972, Abt2002, Preston1974, Smith1996, Mathys1997, Michaud2015, Hummerich2020}. These stars have been shown to be slow rotators with $\vsini<150$\,km/s (Fig.\,\ref{fig:fig16}), which is consistent with previous results \citep[e.g.][]{Abt2002, Preston1974, Michaud2015}. Besides the ApBp stars, our sample may also contain a significant number of other types of chemically peculiar stars, such as Hg-Mn stars, Am/Fm stars, He-deficient stars etc., as our sample stars are uniformly targeted without significant bias against any of these stars. Fig.\,\ref{fig:fig17} shows that at the cool temperature side ($\teff\lesssim10000$\,K), there is a large fraction of stars with super-solar metallicity, forming the high-\feh tail of the \feh distribution. Many of these super-solar metallicity stars are probably AmFm stars, which have been demonstrated to contributed a large fraction ($\sim40$\%) of the A/F stars with mass higher than 1.4\msun in the LAMOST sample \citep{Xiang2020}. 

\begin{figure*}[hbt!]
\centering
\includegraphics[width=1.0\textwidth]{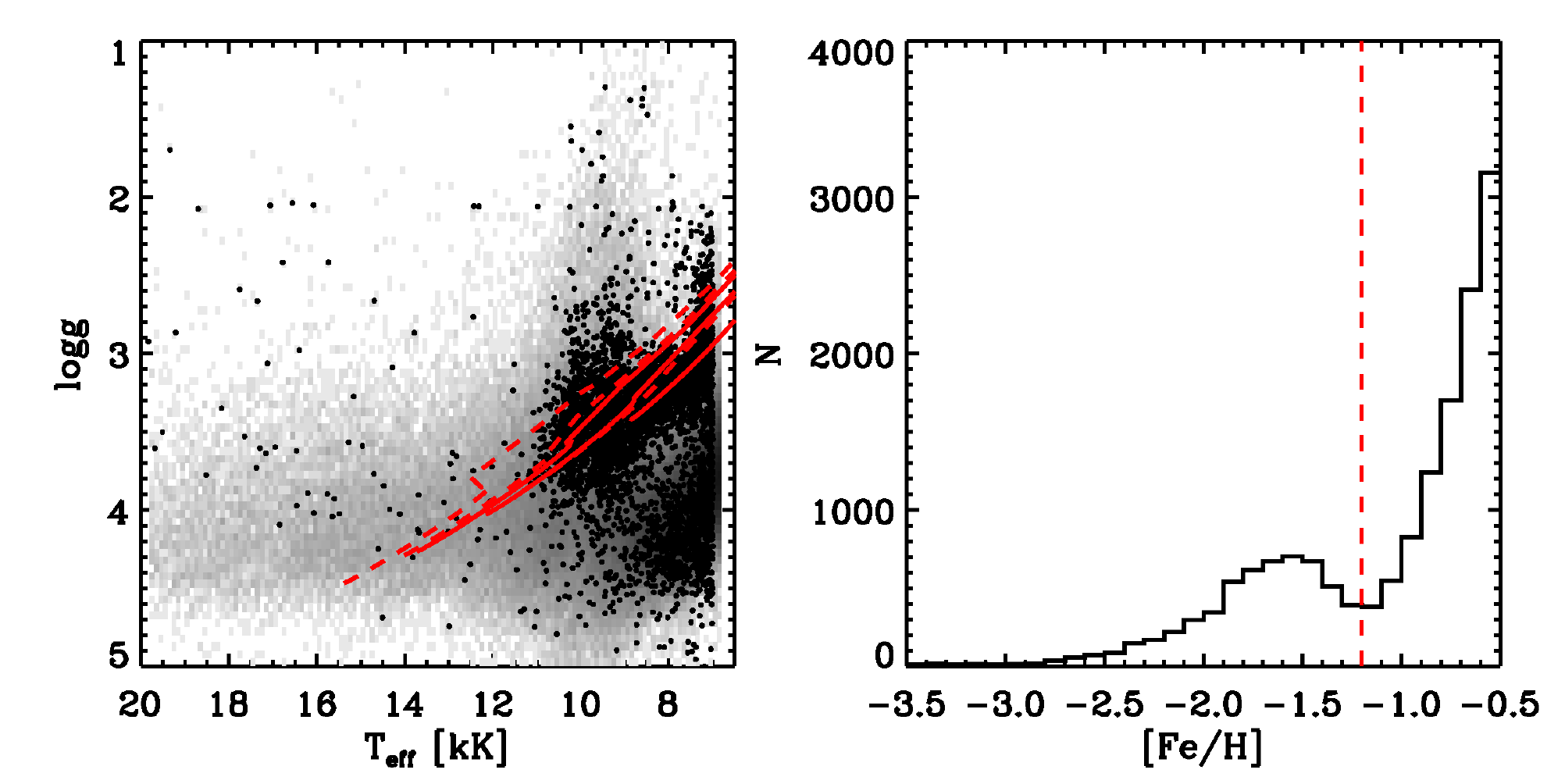}
\caption{Illustration of halo BHB stars revealed in \teff--\logg diagram. Left: the background in grey scale shows the stellar density (in logarithmic scale) of our sample stars with $S/N>50$. The black dots present the stars with $|b|>15^\circ$ and $\feh<-1.2$\,dex. There are two sequences separated in \logg, the upper sequence with lower \logg values is composed of halo BHBs. The solid lines in red are MIST evolutionary tracks of He-burning stars with $\feh=-1.5$, $V/V_{\rm c} = 0.4$, and mass of 0.65, 0.60, 0.58\,$M_\odot$, respectively, from upper to lower. The dash lines in red show similar tracks but for $\feh=-2.0$. Right: [Fe/H] distribution of our sample stars with $S/N>50$ and $|b|>15^\circ$. Only the metal-poor side of $\feh<-0.5$\,dex is shown. The vertical dashed line indicates a constant \feh of $-1.2$\,dex, the sample below which mostly comprises of the metal-poor halo BHB stars.}
\label{fig:fig18}%
\end{figure*}

At the metal-poor side of the \teff--\feh plane, there is a significant number of metal-poor stars with $\feh\lesssim-1.0$\,dex, with a lower \feh border increases from $\sim-2.5$\,dex at $\teff=7000$\,K to $\sim-1.5$\,dex at $\teff=11000$\,K. These metal-poor stars are mostly blue horizontal branch (BHB) stars and blue straggler stars; we expect there are very few main-sequence stars given their high temperatures. As a demonstration, Fig.\,\ref{fig:fig18} shows that stars with $|b|>15^\circ$ exhibit a \feh peak at around $\sim-1.6$\,dex, which is consistent with previous \feh estimates of the nearby halo stars \citep[e.g.][]{Carollo2007, Beers2012, An2013, Zuo2017, Youakim2020}. It is expected that the nearby halo stars dominate our metal-poor sample 
as the majority of them are brighter than 14\,mag (Fig.\,\ref{fig:fig2}). 
Fig.\,\ref{fig:fig18} shows that the BHB stars can be well separated from the blue stragglers in the \teff--\logg diagram. This makes it possible to identify a large set ($\sim5000$) of BHB stars, which have been widely used as tracers to study the structure and dynamics of the Milky Way halo \citep[e.g.][]{Sirko2004a,Sirko2004b, Kinman2007, Xue2008, De_Propris2010, Ruhland2011, Kafle2012, Yang2019, Starkenburg2019, Bird2020}, from the current sample. The BHB stars in our sample can have a temperature higher than 10,000\,K stars, and their distribution in the \teff-\logg diagram is consistent with the prediction of the MIST evolutionary tracks. Whereas, blue stragglers, the product of binary mass transfer, occupy a different locus in the \teff--\logg plane as they have larger \logg values than (and separable from) BHB. Our results also reveal that most of these metal-poor stars are slow rotators with $\vsini\lesssim150$\,km/s (Fig.\,\ref{fig:fig16}). Among them, we find that the BHB stars have a \vsini distribution of $49\pm36$\,km/s, which is qualitatively consistent with literature values \citep{Peterson1995, Cohen1997, Behr2003, Recio-Blanco2004, Cortes2009}, taking into account the relatively large uncertainty in our \vsini estimates.

The right panel of Fig.\,\ref{fig:fig17} illustrates that for the majority of stars of $\teff>8000$\,K, the [Si/H] estimates are consistent with the \feh estimates, while at the metal-poor side (e.g. $\feh<-1.0$\,dex), the [Si/H] is about 0.2\,dex higher than the \feh, which means that the metal-poor halo stars have ${\rm [Si/Fe]}\sim0.2$\,dex. This is consistent with the [$\alpha$/Fe] values of the Galactic halo stars in previous work \citep[e.g.][]{Hayes2018, Conroy2019}. We caution about the [Si/H] estimates for stars with $\teff<8000$\,K, which we find much larger uncertainties, probably due to the weak features of Si in those relatively cool stars.

\subsection{The parameter catalog}
\begin{table*}
\caption{Descriptions for the stellar label catalog for 352,987 hot stars in LAMOST DR6. An electronic version of the catalog is available.}
\label{table:table5}
\begin{tabular}{ll}
\hline
 Field &    Description  \\
\hline
 specid &   LAMOST spectra ID in the format of ``date-planid-spid-fiberid" \\
 fitsname & Name of the LAMOST spectral .FITS file \\
 ra &   Right ascension from the LAMOST DR5 catalog (J2000; deg)\\
 dec &  Declination from the LAMOST DR5 catalog (J2000; deg)  \\
 uniqflag & Flag to indicate repeat visits; uniqflag = 1 means unique star, uniqflag = 2, 3, ..., $n$ indicates the $n$th repeat visit \\
 &  For stars with repeat visits, the uniqflag is sorted by the spectral S/N, with uqflag = 1 having the highest S/N \\
 star\_id  & A unique ID for each unique star based on its RA and Dec, in the format of ``Sdddmmss$\pm$ddmmss" \\
 snr\_g &  Spectral signal-to-noise ratio per pixel in SDSS g-band \\
 rv  & Radial velocity from LAMOST fits header (km/s) \\
 rv\_err &  Uncertainty in radial velocity (km/s) \\
 \teff  & Effective temperature \\
 \teff\_err &  Uncertainty in \teff \\
 \logg  & Surface gravity \\
 \logg\_err &  Uncertainty in \logg \\
 \vsini & Projected rotation velocity \\ 
 \vsini\_err & Uncertainty in \vsini \\
 \feh & Iron abundance \\
 \feh\_err & Uncertainty in \feh  \\
 ${\rm [Si/H]}$ & Silicon abundance \\
 ${\rm [Si/H]}$\_err & Uncertainty in ${\rm [Si/H]}$ \\
 corr\_\teff\_\logg & Correlation coefficient between \teff and \logg, derived from the covariance array of the fit \\
 corr\_\teff\_\vsini  &  Correlation coefficient between \teff and \vsini \\
 corr\_\teff\_\sih  &  Correlation coefficient between \teff and \sih \\
 corr\_\teff\_\feh  &  Correlation coefficient between \teff and \feh \\
 corr\_\logg\_\vsini  &  Correlation coefficient between \logg and \vsini \\
 corr\_\logg\_\sih  &  Correlation coefficient between \logg and \sih \\
 corr\_\logg\_\feh  &  Correlation coefficient between \logg and \feh \\
 corr\_\vsini\_\sih  & Correlation coefficient between \vsini and \sih \\
 corr\_\vsini\_\feh  & Correlation coefficient between \vsini and \feh \\
 corr\_\sih\_\feh  & Correlation coefficient between \sih and \feh \\
 chisq\_red & Reduced $\chi^2$ (i.e., $\chi^2$/dof) of the fit \\
 chi2ratio & Deviation of the chisq\_red from the median value of chisq\_red for stars of similar S/N and \teff, \\
  & divided by the dispersion of the chisq\_red \\
 source\_dr6oba & Target from the LAMOST classification for OBA stars? (1=`YES', 0=`NO') \\ 
 source\_Zari21 & Target from the photometry-based identification of \citet{Zari2021}? (1=`YES', 0=`NO') \\
 source\_Liu19 & Target from the OB star catalog of \citet{LiuZ2019}? (1=`YES', 0=`NO') \\
 gaia\_id & Source ID of Gaia eDR3, matched with the CDS-Xmatch service \footnote{http://cdsxmatch.u- strasbg.fr/xmatch} using 3" criterion  \\
 parallax & Gaia eDR3 parallax \\
 parallax\_error & Uncertainty in parallax \\
 pmra & Gaia eDR3 proper motion in RA direction \\
 pmra\_error & Uncertainty in pmra \\
 pmdec & Gaia eDR3 proper motion in Dec direction \\
 pmdec\_error & Uncertainty in pmdec \\
 RUWE & Gaia eDR3 RUWE \\
 r\_geo & Geometric distance derived from Gaia parallax by \citet{Bailer-Jones2021}. \\
 r\_geo\_low & Distance at 16th percentile of the probability distribution  \\
 r\_geo\_high & Distance at 84th percentile of the probability distribution \\
 fidelity & Value of fidelity from \citet{Rybizki2021}, an identifier for spurious astrometric solutions \\
 ebv\_bayes19 & E(B-V) interpolated from the 3D map of \citet{Green2019} using  r\_geo \\
 G/BP/RP & Gaia eDR3 magnitude  \\
 J/H/K & 2MASS photometric magnitude \\
 \hline
\end{tabular}
\end{table*} 
Our stellar parameter catalog contains \teff, \logg, \vsini, \feh, and [Si/H] for 332,172 unique stars with $\teff>7000$\,K and $S/N>5$ from 454,693 spectra of LAMOST DR6. Table\,\ref{table:table5} presents a description of the columns in our catalog. An electronic version of the catalog is available, either via the cds archive or via the LAMOST website \footnote{http://dr6.lamost.org/v3/doc/vac}. Here we summarize a few notes of caution when using the catalog.
\begin{itemize}
    \item For stars with $\logg>5$, which are mostly hot subdwarfs, the labels should be used with cautious as they are determined with model spectra extrapolated by \hotpayne. 
    \item We assign a \vsini value of 0\,km/s for stars that our measurement gives negative values due to extrapolation. We also caution that although the \vsini estimates should be statistically robust, the uncertainty is large. An inspection of the spectra is necessary when focusing on the \vsini of individual stars, this is particularly true for stars with unexpectedly large \vsini values, as many of them have been found to be artefacts due to either binary effects or data imperfection. 
    \item The labels are deduced based on LTE models, and they may suffer systematics due to NLTE effects as well as stellar wind effects, particularly for stars with $\teff>25,000$\,K, and for stars with low \logg values. 
    \item The label uncertainties presented in the catalog are the formal errors of the fitting, and has been scaled to match the dispersion of repeat observations as a function of \teff and S/N with 3-order polynomial, similar to \citet{Xiang2019}. The uncertainties provided here only concern the internal precision, they do not account for the zero-point offset  induced by imperfections of the adopted model spectra.
\end{itemize}

\subsection{Future improvements}
This is our first attempt of deriving stellar labels for hot stars from a vast set of low-resolution $R\sim1800$ spectra, for which most of the spectral features are very blended and weak, with an automated full-spectral fitting technique. Although we have demonstrated that the $R\sim1800$ optical spectra do contain a wealth of information of hot stars that in principle can give us robust estimates of stellar parameters as well as abundances for many ($>5$) elements, this work comes out with a few imperfections that need to be improved in future work. 
   
First, it is well known that NLTE effects (and wind effects) play an important role for hot-star spectra. In order to obtain accurate labels, particularly for elemental abundances, a large grid of NLTE spectral models are urgently needed. Secondly, it is well known that a large fraction of hot stars are in binary systems \citep[e.g.][]{Sana2012, Sana2013}, future spectroscopy should take the binary effect into account in order to derive reliable label estimates for binary systems. Similarly, many of the LAMOST O- and B-type stars ($>10\%$) have emission lines in their spectra \citep[e.g.][]{Xiang2021}, special care is needed for such systems. Thirdly, in this work we have adopted the mean LSF of the LAMOST spectra, while it is found that the fiber-to-fiber and temporal variations of the LAMOST LSF do have a significant impact on the \vsini estimates. Incorporating the exact LSF for individual stars should improve the \vsini measurements. Finally, we have adopted the stellar line-of-sight velocity from the LAMOST pipeline to correct the spectra into rest frame. However, we do find that, at times, the LAMOST pipeline provides erroneous line-of-sight velocity for some OB stars, particularly those with emission lines in their spectra. In future work, we may derive the stellar labels together with line-of-sight velocity simultaneously from the spectra. We mention that most of the LAMOST OB stars with emission lines in their spectra should have been identified and marked in the catalog of \citet{Xiang2021}. 

\section{Summary}\label{sec:summary}
In this study, we have demonstrated that stellar atmospheric parameters and multiple elemental abundances of hot stars can be deduced from the low-resolution ($R\sim1800$) optical spectra. This is first examined with the Cramer-Rao bound calculation, which suggests that for a typical B star with spectral S/N of $\sim100$, the theoretical limit of the precision one could attain is 300\,K in \teff, 0.03\,cm/s$^2$ in \logg, 1.0\,km/s in \vmic, 10\,km/s in \vsini, 0.01 in \he, $\sim0.1$\,dex in abundances of C, O, Si, S, and Fe, and $\sim0.2$\,dex in abundance of N and Mg. 

We adopt The Payne to derive 11 labels --- \teff, \logg, \vmic, \vsini, \he, [C/H], [N/H], [O/H], [Si/H], [S/H], and [Fe/H] -- from a vast set of LAMOST spectra, taking the LTE-based Kurucz synthetic spectra as the underlying model spectra. Testing with synthetic spectra and examining with LAMOST duplicate observations confirm the robustness of the label estimates. We also show that through the repeated spectra from the same objects that we can attain an internal precision that is within a factor of two from the Cramer Rao bound with real-life LAMOST data. 

Despite the excellent internal precision and consistency, external examinations with literature values suggest that, the abundance estimates may suffer non-negligible zero-point systematics, likely consequences of NLTE effects. Nonetheless, the current estimates of \teff, \logg, and \vsini are likely accurate, except for stars with $\teff>25,000$\,K or supergiants with $\logg\lesssim2$. Considering these non-trivial systematics, we presented the results of five stellar labels, \teff, \logg, \vsini, \sih, and \feh in our catalog, and other stellar labels are available upon request. 

In the \teff--\logg diagram, our LAMOST sample stars cover a broad range of parameter space that corresponds to a stellar mass range from 1.5\,$M_\odot$ to larger than 50\,$M_\odot$. The \vsini distribution decreases roughly with a power law from a peak at $\sim30$\,km/s to $\sim300$\,km/s, above which the stellar density exhibits a cutoff. This terminate \vsini increases with \teff, from $\sim300$\,km/s at 7000\,K to $\sim600$\,km/s at 40,000\,K. This \teff-dependent cutoff is consistent well with the theoretical prediction of the critical rotation velocity. The \vsini distribution also suggests that at all temperature from 7000\,K to 25,000\,K, the stars are dominated by slow rotators with $\vsini<100$\,km/s. 

Our results reveal a sample of ApBp stars with significantly enhanced \feh ($\feh>0.2$\,dex) in the range of $7000<\teff<20,000$\,K, and many of them have also significantly enhanced \sih ($\sih>0.4$\,dex). These ApBp stars have rotation velocity smaller than 150\,km/s, which is consistent with previous studies. The label estimates also allow the identification of a large sample of BHB stars in the \teff--\logg plane. Our catalog of \teff, \logg, \vsini, \sih, and \feh for a sample of 332,172 stars with $\teff>7000$\,K from LAMOST DR6, including more than 64,000 OB stars with $\teff>10,000$\,K, are public accessible. 

Our analysis method and conclusion is generic, and can be applied to other surveys, such as the SDSS-V survey. Future works based on NLTE analysis are desired to improve upon the current results, particularly to obtain accurate stellar abundances from this vast set of LAMOST spectra, as well as the coming SDSS-V spectra.
  
\begin{acknowledgements} 
The authors thank the reviewer, prof. Artemio Herrero, for his careful reading and valuable comments on the manuscript. M. Xiang also thanks Prof. H.-W. Zhang for helpful discussion. YST is grateful to be supported by the NASA Hubble Fellowship grant HST-HF2-51425.001 awarded by the Space Telescope Science Institute.
RPK acknowledges support by the Munich Excellence Cluster Origins funded under germany's Excellence Strategy EXC-2094 390783311. SG gratefully acknowledges support from the Research Foundation Flanders (FWO) by means of a PhD Aspirant mandate under contract No. 11E5620N
The research leading to these results has (partially) received funding from the KU\,Leuven Research Council (grant C16/18/005: PARADISE), from the Research Foundation Flanders (FWO) under grant agreement G0H5416N (ERC Runner Up Project), and from the BELgian federal Science Policy Office (BELSPO) through PRODEX grant PLATO. \\

This work has made use of data products from the Guoshoujing Telescope (the LAMOST). LAMOST is a National Major Scientific Project built by the Chinese Academy of Sciences. Funding for the project has been provided by the National Development and Reform Commission. LAMOST is operated and managed by the National Astronomical Observatories, Chinese Academy of Sciences. \\

This work has made use of data products from the European Space Agency (ESA) space mission Gaia. Gaia data are being processed by the Gaia Data Processing and Analysis Consortium (DPAC). Funding for the DPAC is provided by national institutions, in particular the institutions participating in the Gaia MultiLateral Agreement. The Gaia mission website is https://www.cosmos.esa.int/gaia. The Gaia archive website is https://archives.esac.esa.int/gaia.\\
 
This publication has also made use of data products from the 2MASS, which is a joint project of the University of Massachusetts and the Infrared Processing and Analysis Center/California Institute of Technology, funded by the National Aeronautics and Space Administration and the National Science Foundation. 

This research has made use of the TOPCAT software, the IDL software, and the Python software. \\

\end{acknowledgements}

%
   \bibliographystyle{aa} 
   \bibliography{refer} 
%



\appendix
\section{The gradient spectra of early A-type stars}
\begin{figure*}[hbt!]
 \centering
 \includegraphics[width=1.0\textwidth]{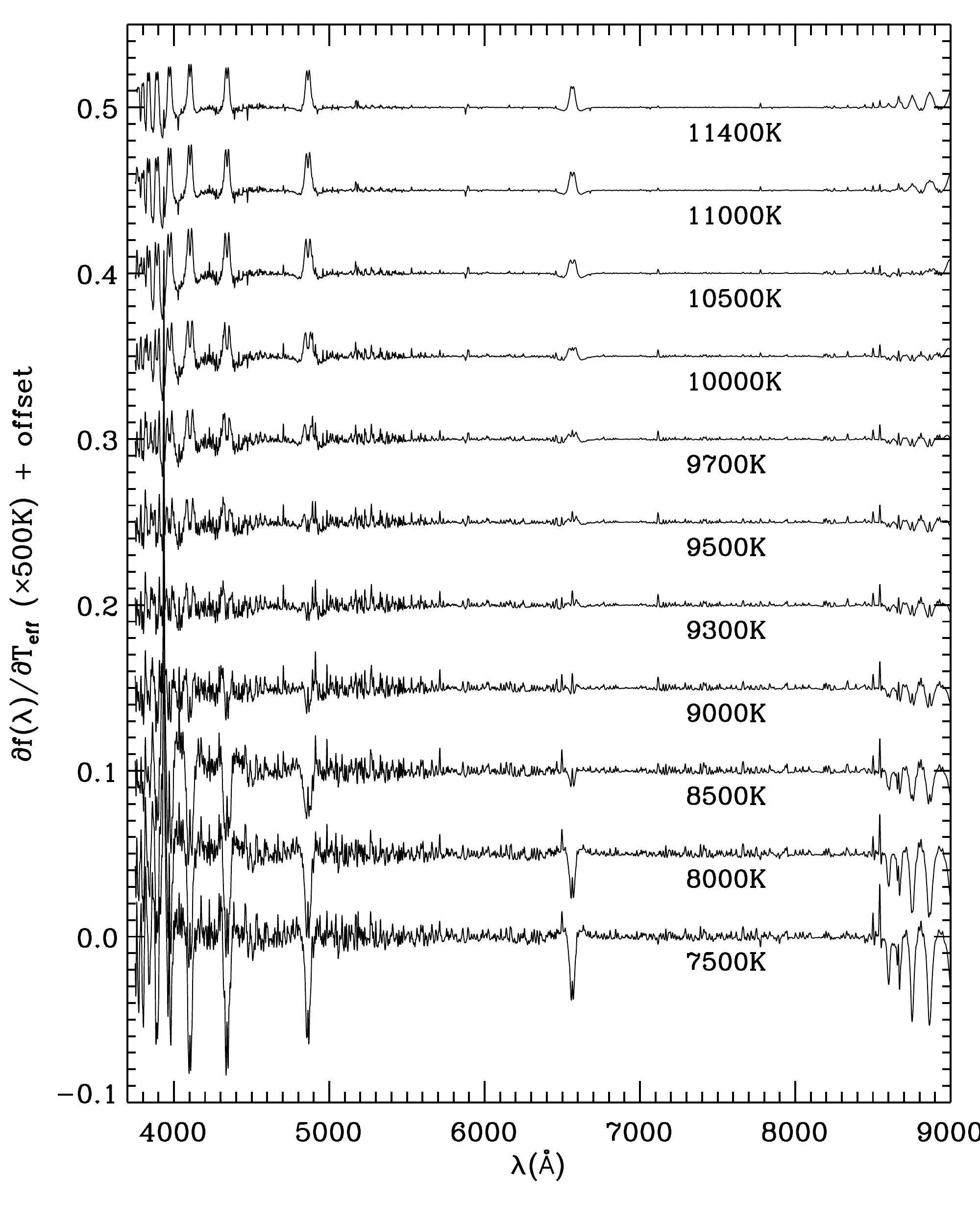}
 \caption{Gradient spectra of \teff derived from the Kurucz model at different effective temperatures, as marked in the figure. We assume \logg=4.5, and \feh=0. The A0 dwarf ($\teff=9500$\,K) has the minimal (close to 0) gradients in Hydrogen lines, and they form an inflection point in the  gradient of the Hydrogen lines as a function of \teff, which leads to the artificial A0 star desert of the stellar distribution in the \teff-\logg diagram as seen in Fig.\,\ref{fig:fig9}.  }
 \label{fig:figA1}%
\end{figure*}

The A0-type dwarfs have the strongest Hydrogen lines in their spectra, which means that the gradient spectra $\partial{f}_\lambda/\partial\lambda$ of Hydrogen lines reach a minimum for these stars. As an illustration, Fig.\,\ref{fig:figA1} shows the gradient spectra of dwarf stars ($\logg=4.5$, $\feh=0$) at different temperatures. The gradient spectra of Hydrogen lines have negative values at $\teff<9500$\,K, and become positive values at $\teff>9500$\,K. This reversal leads to the A0 ``desert'' shown in Fig.\,\ref{fig:fig9}.

\section{The Cram\'er-Rao bound of A stars}
\begin{figure*}[hbt!]
\centering
 \includegraphics[width=0.8\textwidth]{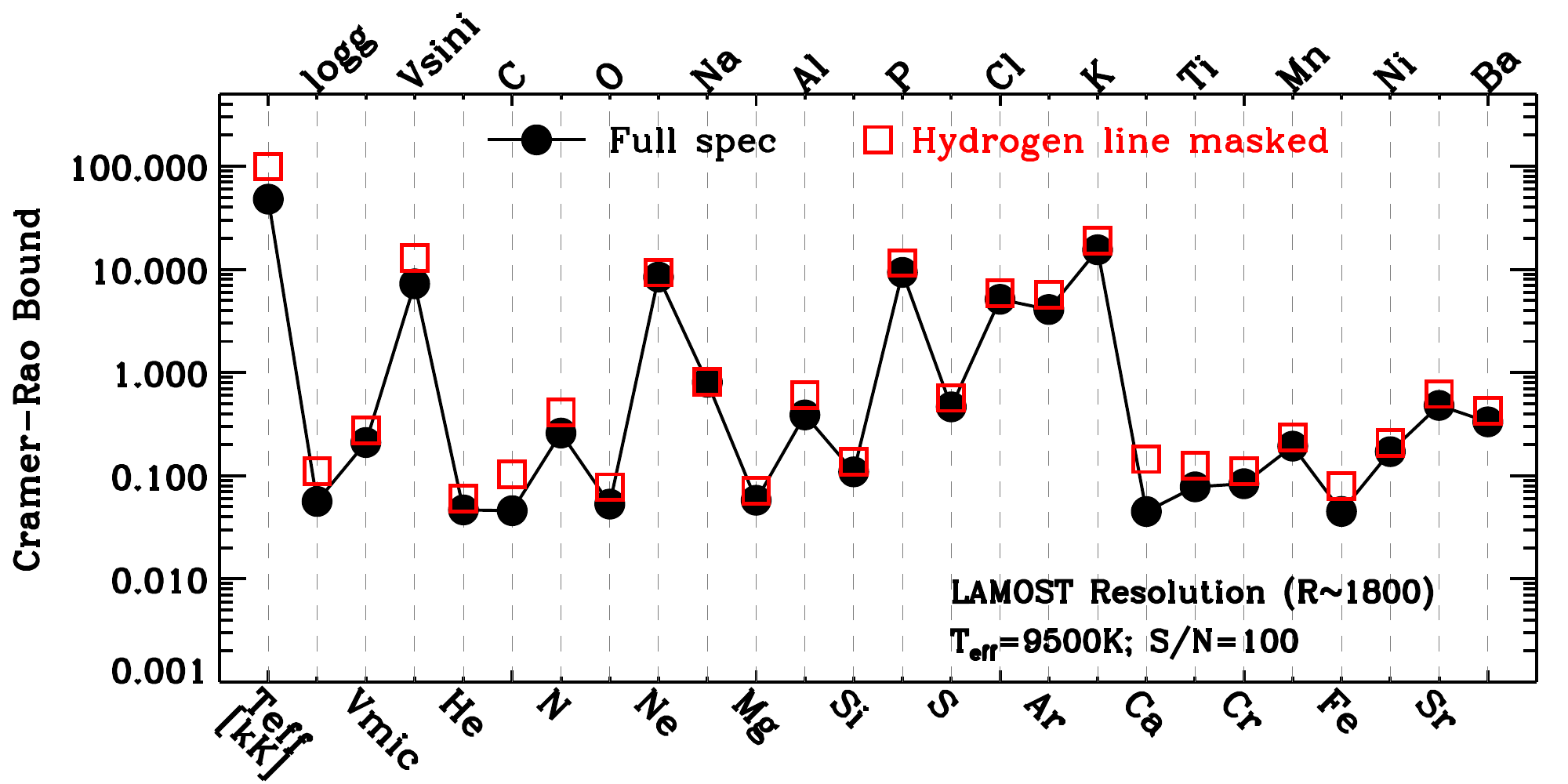}
 \caption{CR bound for an A-type star with $\teff=9500$\,K, $\logg=4.5$, and $\feh=0$. The black dots show the CR bound derived from the full spectra, and the red squares are derived from the spectra after masking the Hydrogen lines as described in Section\,4.5.}
 \label{fig:figA2}%
\end{figure*}
Fig.\,\ref{fig:figA2} shows the Cram\'er-Rao bound for an A-type dwarf star with $\teff=9500$\,K, $\logg$=4, and $\feh=0$, both from the full spectra and from the spectra after masking Hydrogen lines as introduced in Section\,4.5. Masking Hydrogen lines loses a part of the astrophysical information content, leading to larger CR bound. Nonetheless, the values of the CR bound i.e., theoretical precision limit of the label estimates, are still feasible, with a precision of 100\,K in \teff, 0.1\,dex in \logg, 0.3\,km/s in \vmic, 15\,km/s in \vsini, and 0.1\,dex in the abundance for a few elements (e.g. C, O, Mg, Fe) for a spectral S/N of 100.

\section{Comparison between LTE and NLTE model spectra}
\begin{figure*}[hbt!]
\centering
\includegraphics[width=0.9\textwidth]{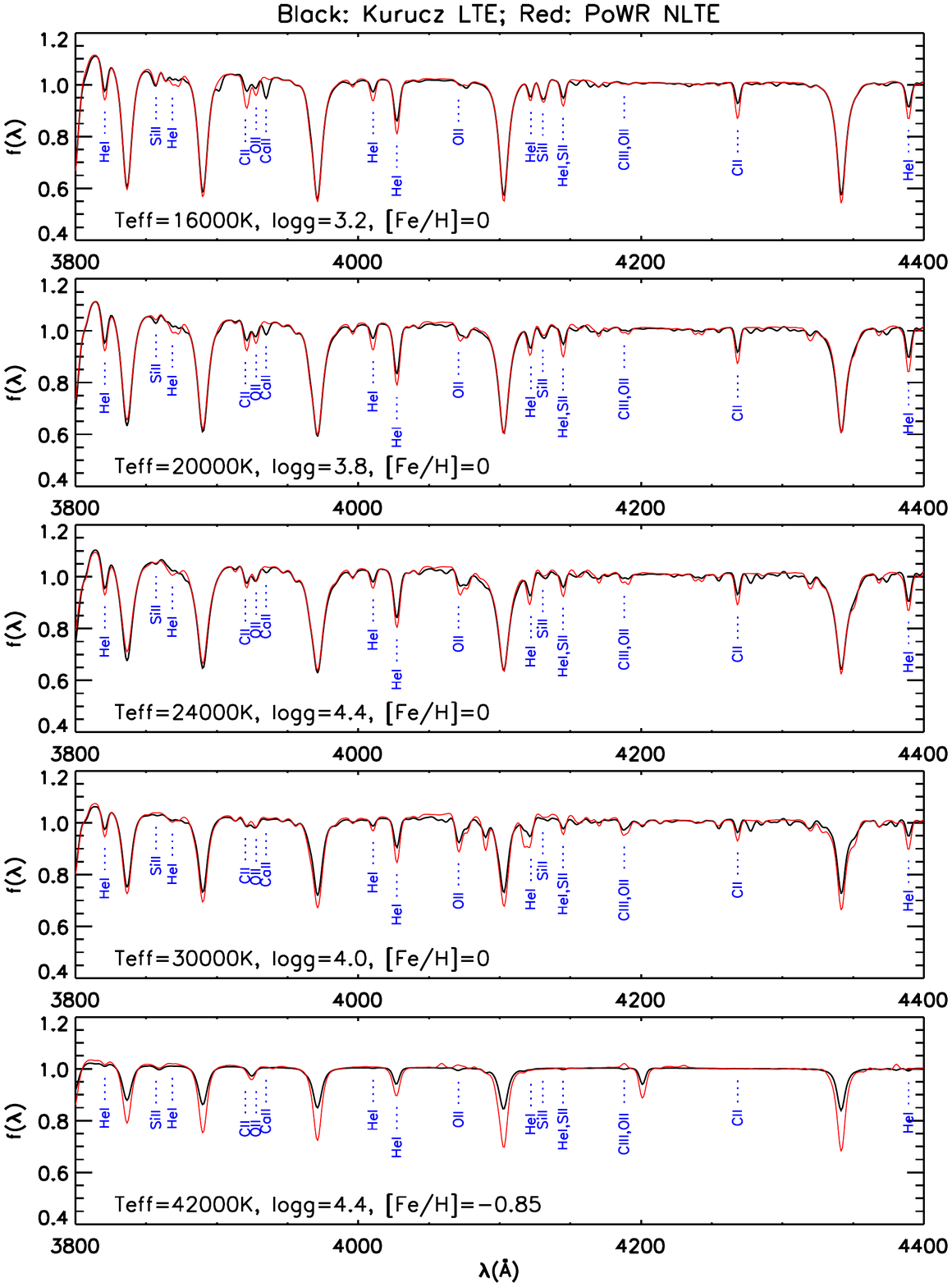}
\caption{Comparison of Kurucz LTE model spectra (black) with the PoWR NLTE model spectra (red) for different parameters. The NLTE effect has a strong impact on the He lines and metal lines at all temperatures, and have also strong impact on the Hydrogen lines for stars with $\teff\gtrsim25,000$\,K. }
\label{fig:figA3}%
\end{figure*}

\begin{figure*}[hbt!]
\centering
\includegraphics[width=0.9\textwidth]{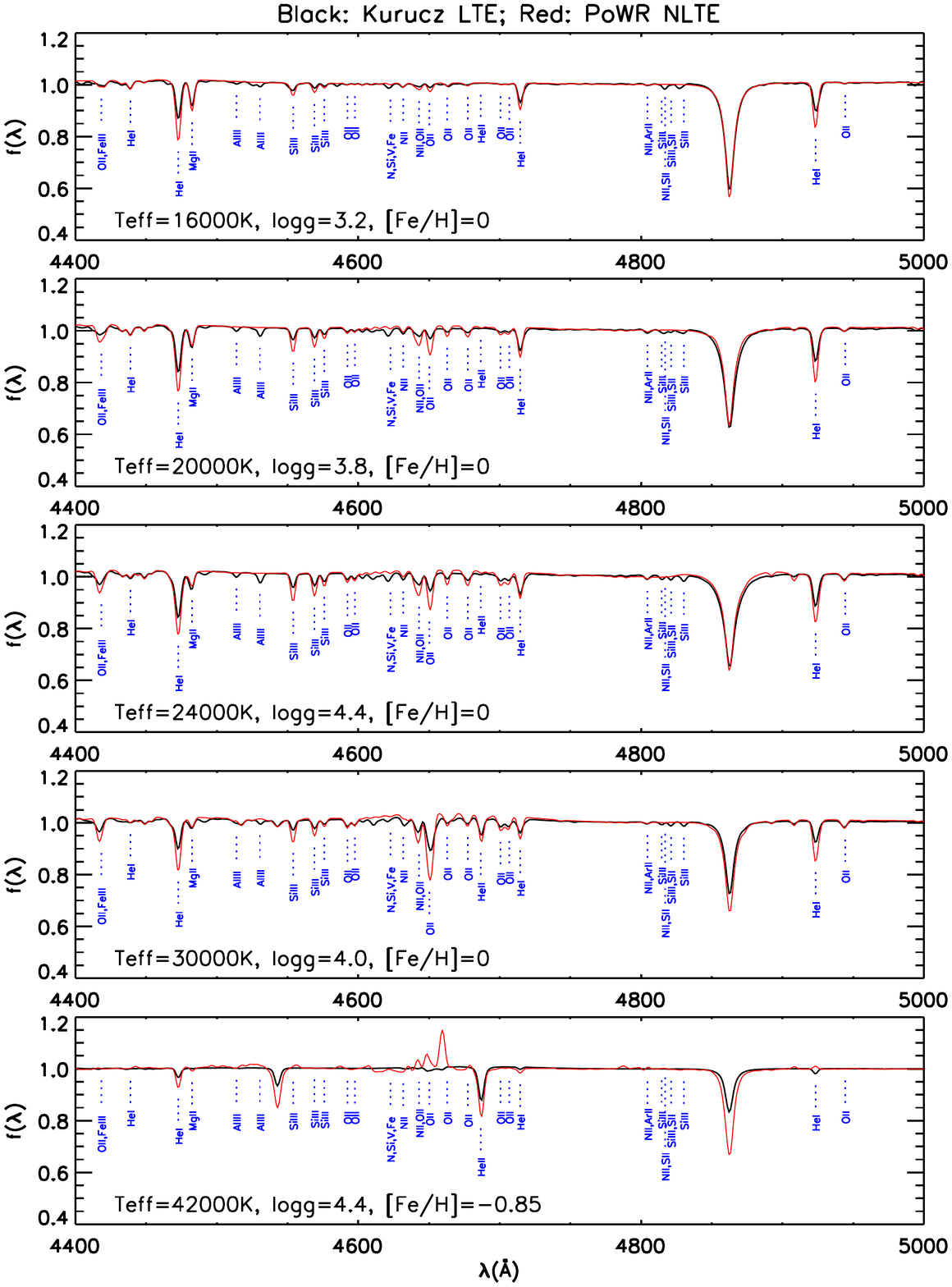}
\caption{Same as Fig.\,\ref{fig:figA3}, but for the wavelength range $\lambda$4400--5500{\AA}.}
\label{fig:figA4}%
\end{figure*}
Figs.\,\ref{fig:figA3} and \ref{fig:figA4} present a comparison of the Kurucz LTE model spectra with NLTE spectra from the Potsdam Wolf-Rayet Models \citep[PoWR;][]{Hainich2019}, for different temperatures. The NLTE (and stellar winds) effect has a strong impact on the He lines at all temperatures, and have strong impact on the Hydrogen lines for stars with $\teff\gtrsim25,000$\,K \citep[see also e.g.][]{Auer1972, Kudritzki1976, Kudritzki1979}. 

\end{document}